\documentclass{article}

\usepackage{arxiv}
\usepackage[utf8]{inputenc}

\usepackage{amsmath,amssymb,amsfonts}
\usepackage{nicefrac}

\usepackage{graphicx}
\usepackage{float}
\usepackage{subcaption}

\usepackage{booktabs}
\usepackage{multirow}
\usepackage{makecell}

\usepackage{textcomp}
\usepackage{microtype}
\usepackage{url}
\usepackage{xcolor}
\usepackage{pifont}
\usepackage[nointegrals]{wasysym}

\usepackage{algorithmic}
\usepackage{listings}
\usepackage{tcolorbox}
\tcbuselibrary{listings,breakable,skins}

\usepackage{hyperref}
\usepackage{cleveref}

\usepackage{changepage}

\usepackage[T1]{fontenc}
\usepackage{lmodern}
\usepackage{anyfontsize}

\definecolor{myblue}{RGB}{30,70,120}
\definecolor{mycomment}{RGB}{120,120,120}

\lstdefinestyle{pseudo}{
  frame=single,
  numbers=left,
  numberstyle=\small,
  stepnumber=1,
  numbersep=5pt,
  xleftmargin=15pt,
  framexleftmargin=15pt,
  basicstyle=\ttfamily\small,
  keywordstyle=\bfseries,
  commentstyle=\color{gray!70},
  showstringspaces=false,
  tabsize=4,
  breaklines=true,
  captionpos=t,
  literate=
    {Δ}{{$\Delta$}}1
    {α}{{$\alpha$}}1
}

\usepackage{changepage}

\newcommand{\estimationheading}[1]{%
  \par
  \addvspace{0.75\baselineskip}%
  \noindent\textbf{#1}\par
  \nobreak
}

\usepackage{hyperref}

\usepackage{cleveref}

\title{Flex-Plorer: A Precision-Aware Hardware--Software
       Co-Design Framework for an Edge SNN Accelerator}

\date{}

\author{ Mohammad Farahani~\textsuperscript{\href{https://orcid.org/0009-0001-4267-6557}{\includegraphics[width=8pt]{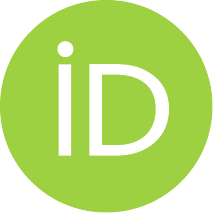}}}\\
	School of Electrical and Computer Engineering\\
	College of Engineering, University of Tehran\\
	Tehran, Iran \\
	\texttt{farahani.mohammad@ut.ac.ir} \\
	\And
	Mohammad~Rasoul Roshanshah~\textsuperscript{\href{https://orcid.org/0000-0001-5650-7925}{\includegraphics[width=8pt]{orcid.pdf}}}\\
	School of Electrical and Computer Engineering\\
	College of Engineering, University of Tehran\\
	Tehran, Iran \\
	\texttt{mrroshanshah@ut.ac.ir} \\
	\And
	Saeed Safari~\textsuperscript{\href{https://orcid.org/0000-0001-6940-591X}{\includegraphics[width=8pt]{orcid.pdf}}}\\
	School of Electrical and Computer Engineering\\
	College of Engineering, University of Tehran\\
	Tehran, Iran \\
	\texttt{saeed@ut.ac.ir} \\
}

\renewcommand{\shorttitle}{Flex-Plorer}

\hypersetup{
	pdftitle={Flex-Plorer}
}

\begin{document}

\maketitle

\begin{abstract}
  Neuromorphic accelerators promise high energy efficiency and computational
  density for spiking neural networks~(SNNs), especially in wearable devices and
  neural prosthetics where power constraints are stringent. However, most existing
  platforms exhibit rigid architectures with limited configurability, which
  restricts their adaptability to heterogeneous signals and diverse design
  objectives.  To address this gap, we present Flex-Plorer, a precision-aware
  hardware--software co-design framework using heuristic search
  over weight and decay precisions for a fixed architecture.
  It couples hardware-aware fixed-point simulation with
  synthesis-trained LUT/FF estimators and analytical BRAM estimation
  to guide user-defined accuracy--resource trade-offs without
  per-candidate synthesis. The best-found configuration generates
  specialized RTL, drivers, and deployment data.
  Its substrate, Flexi-NeurA, supports multiple neuron models and
  topologies through event-driven, time-multiplexed cores with
  shared neuron datapaths and address-event communication.
  Design-time component selection and automatic memory sizing
  specialize the implementation without manual RTL modification.
  Comprehensive evaluations demonstrate the versatility of the proposed flow
  across three application domains: (i)~auditory processing, achieving $75\%$ 
  accuracy with an energy consumption of $1.48~\mathrm{mJ}$ per inference; (ii)~dynamic vision sensor gesture
  recognition, achieving $93\%$ accuracy at $0.55~\mathrm{mJ}$; and
  (iii)~standard vision classification, achieving $97\%$ accuracy at
  $0.12~\mathrm{mJ}$. Together, Flex-Plorer and Flexi-NeurA form an
  application-adaptive, energy-efficient co-design flow for deploying SNNs across
  heterogeneous edge-AI workloads.
\end{abstract}

\keywords{
  Neuromorphic computing \and
  Spiking neural networks \and
  Configurable hardware \and
  Edge computing \and
  Precision optimization \and
  Design-space exploration
}

\section{Introduction}\label{sec:introduction}

Edge-AI refers to executing AI models at or near the location where data is generated, rather than relying solely on distant cloud infrastructure \cite{ref1}. Physical proximity to the data source yields the properties edge systems value most: reduced latency, lower bandwidth consumption, and improved privacy \cite{ref2}.

Spiking neural networks (SNNs) represent information through discrete spike events and evolving neuronal states \cite{ref3,ref4,ref5,ref26}. Their sparse activity can reduce synaptic computation and data movement, making them a promising candidate for energy-efficient edge inference \cite{ref7}. Realizing these benefits, however, depends on how neuron dynamics, and numerical precision are mapped onto hardware. Because edge devices operate under tight resource and energy budgets, SNN deployment must balance inference accuracy against hardware resource usage, energy consumption, and latency \cite{ref27,ref33}.

Hardware--software co-design connects network-level choices to their
implementation costs~\cite{ref36}, as illustrated by neuromorphic
frameworks such as SpikeFlow~\cite{ref34}, Xiao~et~al.~\cite{ref35},
and SpikeExplorer~\cite{ref12}. For a configurable SNN accelerator,
precision selection requires evaluating both quantized inference
accuracy and the resources required by the corresponding hardware.
Accuracy evaluation must account for the accelerator's arithmetic
and neuron-update order, while resource evaluation must reflect its
datapath and memory organization. Synthesizing every candidate is
costly; evaluating quantization without these implementation details
can obscure the accuracy--resource trade-off.

To address this challenge, we introduce \mbox{Flex-Plorer}
(\underline{Flex}ible Ex\underline{plorer}), an end-to-end
hardware/software co-design framework that automates the flow from
architectural specification to synthesizable RTL for Flexi-NeurA.
Through a graphical interface, the designer fixes all architectural
and network parameters except four precision variables: feed-forward
weight bit-width, recurrent weight bit-width, and the two decay
precisions. The designer defines their search ranges
together with the relative weights assigned to hardware cost and
hardware-aware inference accuracy.
\mbox{Flex-Plorer} then explores this space using two coupled operators.
A \emph{Hardware Estimator} predicts LUT and FF utilization using
synthesis-trained ML models and BRAM utilization using an
analytical model, avoiding per-candidate FPGA synthesis.
A \emph{Hardware- and Dataflow-Aware Accuracy Simulator} evaluates
the quantized network using fixed-point arithmetic and the hardware's
neuron-update order. Just-in-time compilation enables fast simulation
while preserving all modeled arithmetic and dataflow details.
The resource and accuracy estimates are merged into a single
user-weighted cost, and a heuristic search seeks to minimize
this cost over the four precision variables while keeping all other
parameters fixed. The lowest-cost configuration found during the
search is then forwarded automatically to the \emph{RTL Configurator},
which emits synthesizable RTL, drivers, and encoded deployment data.

\begin{figure}[!ht]
  \centering
  \includegraphics[width=\linewidth]{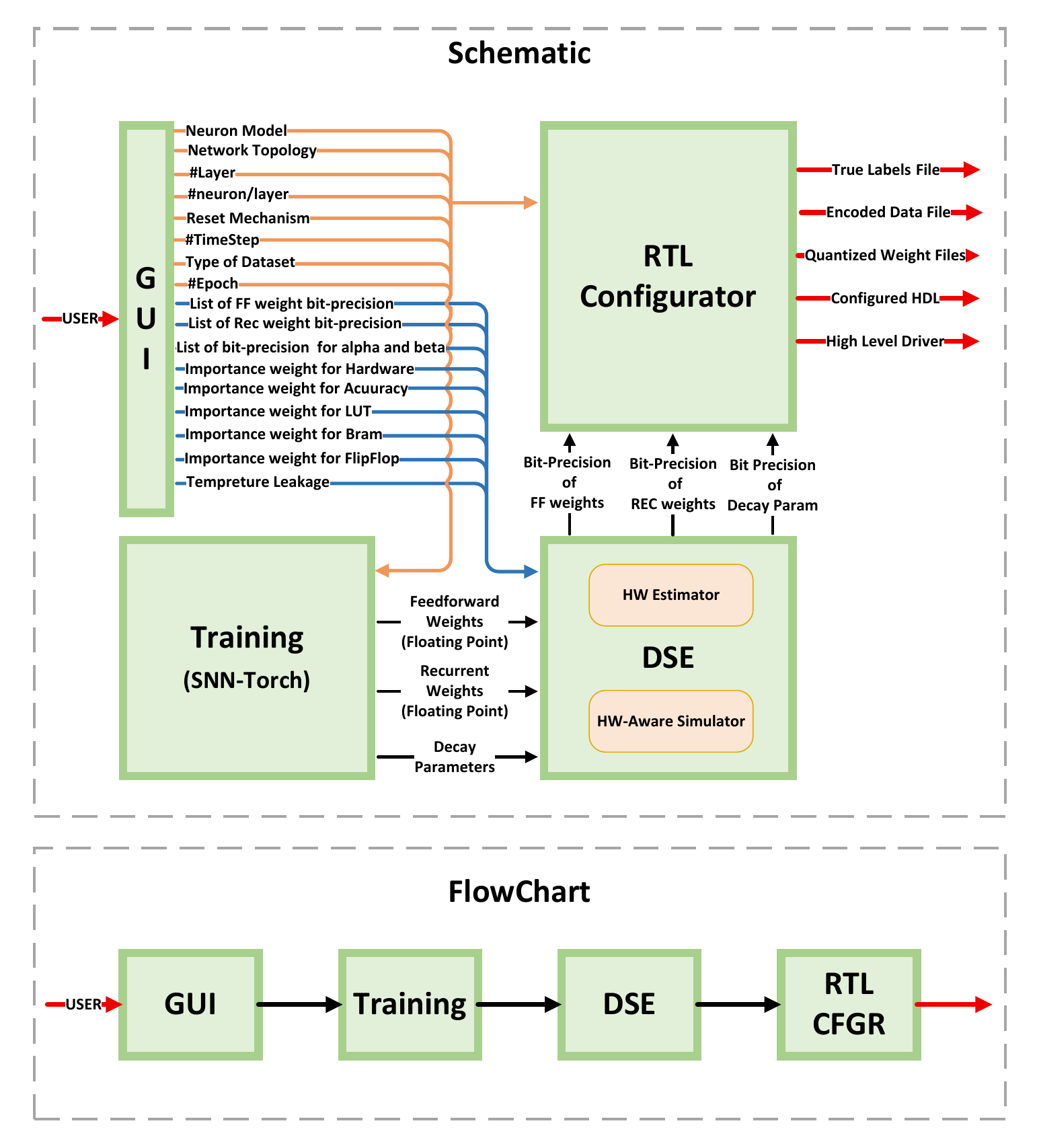}
  \caption{Flex-Plorer workflow linking hardware-aware precision
exploration to configuration-specific RTL and deployment artifacts.}

  \label{fig:Flex-Plorer}
\end{figure}

Event-driven processing, time-multiplexing, multi-core-compatible organization, and design-time configurability constitute fundamental design principles in neuromorphic hardware, each trading energy, area, throughput, and flexibility differently. Event-driven hardware skips computation for inactive neurons, reducing data movement, latency, and power. Time-multiplexing shares computational resources across neurons to minimize the hardware footprint, at the cost of higher latency, whereas multi-core organization improves throughput and latency at the expense of area and power. Configurability, in turn, allows a single substrate to serve multiple workloads, but adds control and routing overhead. Existing accelerators---\mbox{ReckOn}~\cite{ref10}, \mbox{ODIN}~\cite{ref9}, \mbox{MorphIC}~\cite{ref11}, \mbox{Spiker+}~\cite{ref8}, Li~et~al.~\cite{ref22}, Han~et~al.~\cite{ref21}, SpikeFlow~\cite{ref34}, and Xiao~et~al.~\cite{ref35}---adopt these principles selectively according to their design objectives, as summarized in Table~\ref{tab:accelerator_comparison}. Edge and wearable systems, however, require a careful balance among all of them under stringent energy and resource constraints.

We therefore introduce Flexi-\allowbreak NeurA
(\underline{Flexi}ble \underline{Neur}omorphic \underline{A}ccelerator).
This parameterized, multi-core-compatible architecture supports IF, LIF,
and Synaptic neuron models with feedforward and recurrent
connectivity. Each core time-multiplexes a neuron-compute
datapath across its assigned neurons, while address-event
communication connects the cores. At design time, model-
and topology-dependent components are selectively instantiated,
and synaptic and neuron-state memories are sized according
to network dimensions and numerical precisions. This
organization supports configuration-specific implementations
within a common architecture, without manual RTL modification.

\begin{table}[t]
  \centering
  \caption{Comparison of hardware SNN accelerators.}
  \label{tab:accelerator_comparison}
  \footnotesize
  \setlength{\tabcolsep}{4pt}
  \renewcommand{\arraystretch}{1.2}
  \begin{tabular}{@{}clcccccc@{}}
    \toprule
    \textbf{\#} & \textbf{Accelerator} & \textbf{Venue} & \textbf{MC/PE} & \textbf{ED} & \textbf{TM} & \textbf{DTC} & \textbf{CD} \\
    \midrule
    1 & ODIN~\cite{ref9}        & TBioCAS'19 & - & \checkmark & \checkmark & - & - \\
    2 & MorphIC~\cite{ref11}     & TBioCAS'19 & \checkmark & \checkmark & \checkmark & - & - \\
    3 & Han et al.~\cite{ref21}  & TST'20     & - & \checkmark & - & - & - \\
    4 & Li et al.~\cite{ref22}   & TCAS-I'21  & \checkmark & \checkmark & - & - & - \\
    5 & ReckOn~\cite{ref10}      & ISSCC'22   & - & \checkmark & \checkmark & - & - \\
    6 & Spiker+~\cite{ref8}     & TETC'24    & \checkmark & - & - & \checkmark & \checkmark \\
    7 & SpikeFlow~\cite{ref34}   & JSA'25     & \checkmark & - & \checkmark & - & \checkmark \\
    8 & Xiao et al.~\cite{ref35} & JSA'25     & \checkmark & \checkmark & - & - & \checkmark \\
    \midrule
    9 & \textbf{Flexi-NeurA} & \textbf{This work} & \textbf{\checkmark} & \textbf{\checkmark} & \textbf{\checkmark} & \textbf{\checkmark} & \textbf{\checkmark} \\
    \bottomrule
  \end{tabular}

  \vspace{4pt}
  {\scriptsize
    \textbf{MC/PE}: Multi-core / multi-PE parallelism \quad
    \textbf{ED}: Event-driven \quad
    \textbf{TM}: Time-multiplexing \\
    \textbf{DTC}: Design-time configurability \quad
    \textbf{CD}: Hardware--software co-design
  }
\end{table}

The main contributions of this work are summarized as follows:

\begin{itemize}
  \item \textbf{Heterogeneous Precision Co-Design:}
  To overcome the fixed-precision limitations of existing SNN frameworks, we introduce an automated framework that optimizes heterogeneous quantization across feedforward weights, recurrent weights, and neuronal leakages. Guided by a user-weighted accuracy--resource objective, Flex-Plorer searches this multi-dimensional precision space and automatically generates configuration-specific RTL, software drivers, and deployment data.

  \item \textbf{Synthesis-Free, Hardware-Aware Evaluation and Heuristic Search:}
  To evaluate candidate configurations rapidly without per-candidate FPGA synthesis, we couple a dataflow-aware fixed-point simulator with a hybrid hardware resource estimator. This estimator combines synthesis-trained regression models for non-linear logic scaling (LUTs and flip-flops) with a dynamic-programming memory allocator that captures discrete BRAM primitive packing. We further accelerate exploration through a heuristic search algorithm.

  \item \textbf{Configurable Event-Driven Neuromorphic Architecture:}
  We propose an event-driven, time-multiplexed SNN substrate that specializes IF, LIF, and Synaptic neuron models across feedforward and recurrent topologies at design time. Its microarchitecture introduces three key mechanisms: (i) a multiplierless, prunable shift-and-add Coefficient Generator that approximates state decay; (ii) an Address-Event Representation (AER) protocol that helps establish compatibility with multi-core platforms; and (iii) automated memory sizing and byte-aligned banking with automatic RTL generation.
\end{itemize}

The remainder of this paper is organized as follows.
Section~\ref{sec:background} introduces SNN fundamentals and the neuron models considered in this work.
Section~\ref{sec:related_work} reviews existing SNN accelerators and hardware--software co-design frameworks.
Section~\ref{sec:Flexi-NeurA} presents the Flexi-NeurA architecture, its configurable processing unit, communication interfaces, and control mechanisms.
Section~\ref{sec:Flex-Plorer} describes the Flex-Plorer framework, including hardware- and dataflow-aware accuracy simulator, hardware resource estimation, precision-aware design-space exploration, and automated RTL and driver generation.
Section~\ref{sec:results} evaluates the proposed framework on auditory, gesture, and image classification benchmarks, presents quantization exploration results, and compares Flexi-NeurA with state-of-the-art FPGA-based accelerators.
Finally, Section~\ref{sec:conclusion} summarizes the main contributions and findings.

\section{Background}\label{sec:background}

This section reviews the essential SNN background needed for the rest of the paper.

\subsection{Spiking Neural Networks}

SNNs encode information as discrete spikes rather than continuous activations, so computation is event- and time-based, driven by the timing of the spikes. Implementations range from analog to digital; here, we focus on digital SNNs, where a spike is a binary event (1/0) and neurons act as compact modules exchanging these event streams.

Neuronal connectivity in SNNs spans a wide range of architectures, from reservoir networks with irregular, probabilistic synaptic connections~\cite{ref13}, to structured recurrent networks that follow regular connection patterns.

\begin{figure}[t]
  \centering
  \includegraphics[width=\linewidth]{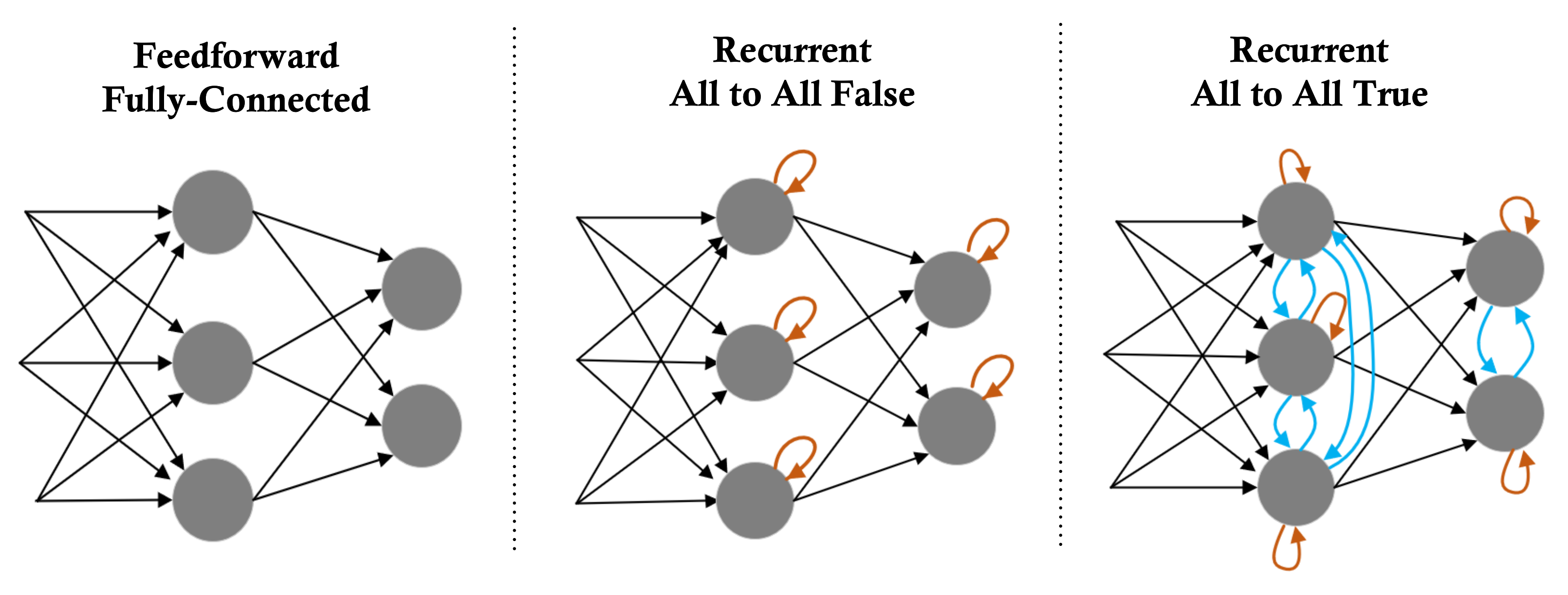}
  \caption{Spiking neural network topologies: (A) fully connected feedforward, (B) recurrent all-to-all false, and (C) recurrent all-to-all true.}
  \label{fig:topologies}
\end{figure}

Within the range of spiking network topologies, Flexi-NeurA supports three implementations:

\paragraph{Fully Connected Feedforward (FF)}
Every neuron of a layer is connected through synapses to all neurons of the
immediately preceding layer (Fig.~\ref{fig:topologies}A).

\paragraph{Recurrent All-to-All False (ATA-F)}
This topology introduces temporal memory at low cost by restricting recurrence to
self-feedback (Fig.~\ref{fig:topologies}B). Each neuron receives feedforward spikes
from all neurons of the previous layer together with its own output from the prior
time step, so that complexity and power consumption remain below those of fully
recurrent designs.

\paragraph{Recurrent All-to-All True (ATA-T)}
This topology strengthens temporal memory through dense intra-layer feedback
(Fig.~\ref{fig:topologies}C). At each time step, every neuron receives feedforward
spikes from all neurons of the preceding layer, as well as recurrent inputs from the
previous time step comprising its own output and the outputs of all neurons within
the same layer.

\subsection{Neuron Models}

Neuron models span a spectrum from the biophysically accurate but computationally costly Hodgkin--Huxley model~\cite{ref15,ref16,ref17} to the reduced-complexity Izhikevich model. For edge hardware, LIF and its variants (IF, synaptic) offer the best trade-off, preserving sufficient biological plausibility while enabling compact, low-power implementations~\cite{ref18}.

\begin{figure}[t]
  \centering
  \includegraphics[width=\linewidth]{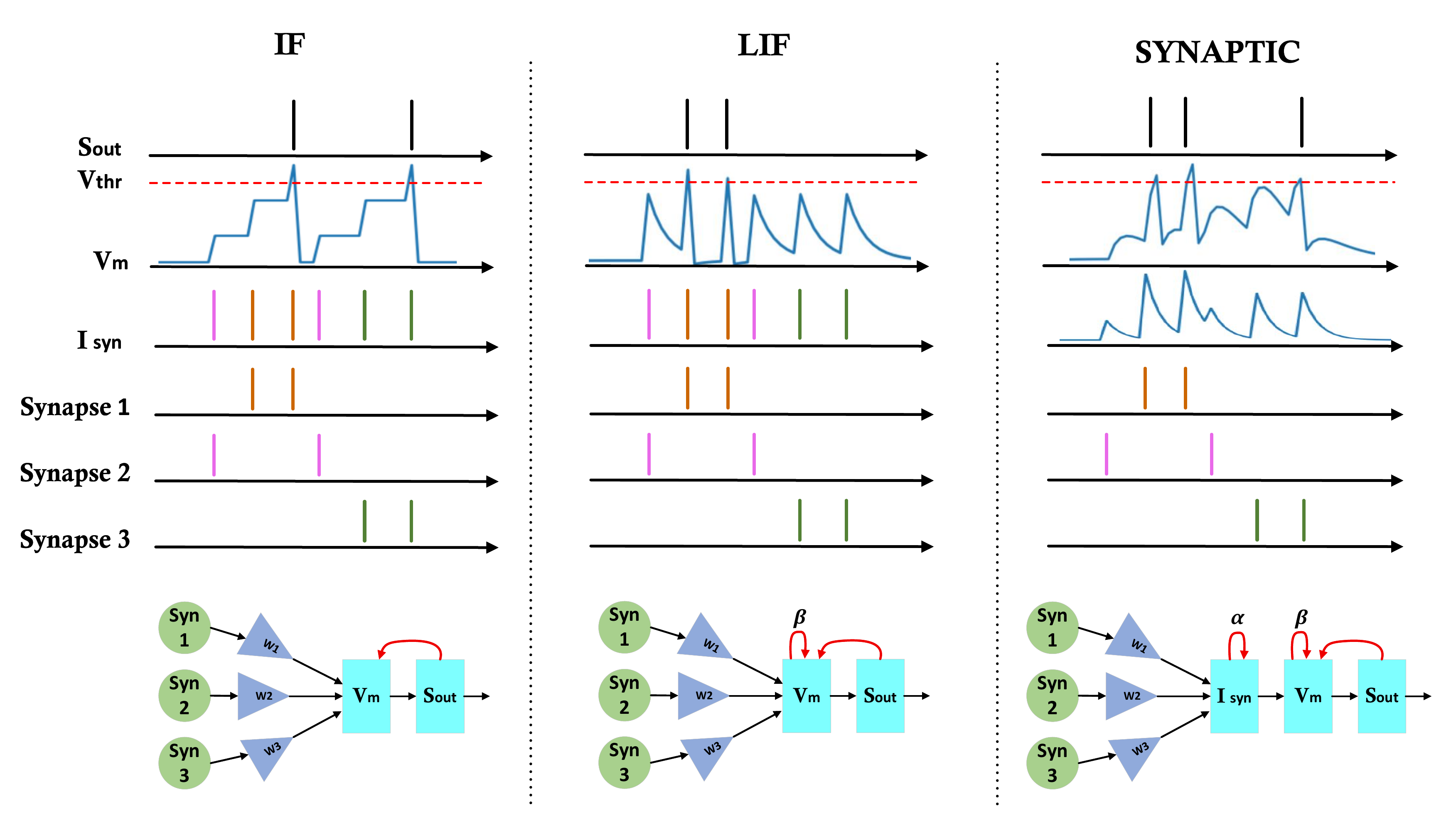}
  \caption{Spiking neuron models and their dynamic responses: (A) Integrate-and-Fire (IF), (B) Leaky Integrate-and-Fire (LIF), and (C) Synaptic neuron.}
  \label{fig:neurons}
\end{figure}

Before formally defining the neuron models, we introduce the notation. The input current $I_{\mathrm{in}}[t]$ at each time step is the weighted sum of incoming spikes, which---depending on network topology---originate from the preceding layer, self-feedback, or lateral (same-layer) connections at the previous time step. Temporal integration is performed on the synaptic current $I_{\mathrm{syn}}[t]$, the membrane potential $U[t]$, or both, depending on the specific neuron model. A spike is emitted ($S[t+1]=1$) once $U[t]$ exceeds the threshold $U_{\mathrm{thr}}$, with the event registered at the subsequent time step. In the absence of a spike, both membrane and synaptic states undergo leakage at every step, governed by retention factors $\beta$ and $\alpha$ ($\alpha,\beta\in(0,1)$). When a spike occurs, however, the membrane potential is reset via the reset operator $R$. As shown in Fig.~\ref{fig:neurons}, each model receives synapses driven by the spike patterns, facilitating a visual comparison of their dynamics.

\paragraph{IF Model} The Integrate-and-Fire (IF) neuron, which operates without leakage, updates its membrane potential by adding the current step's input current to the previous potential and subtracting the reset term upon spike occurrence:
\begin{equation}
  U[t+1] = U[t] + I_{\mathrm{in}}[t+1] - R
  \label{eq:1}
\end{equation}

\paragraph{LIF Model} In this model (Eq.~\ref{eq:2}), the input current $I_{\mathrm{in}}$ is the weighted sum of presynaptic spikes. As depicted in Fig.~\ref{fig:neurons}B, the membrane potential $U$ combines this $I_{\mathrm{in}}$ with a leaky carryover of the previous $U$; if the result exceeds $U_{\mathrm{thr}}$, the neuron fires and the reset rule is applied. Otherwise, $U$ decays by the leak factor $\beta$ and is carried to the next step:
\begin{equation}
  U[t+1] = \beta U[t] + I_{\mathrm{in}}[t+1] - R
  \label{eq:2}
\end{equation}

\paragraph{Synaptic Model} In the synaptic neuron model (Eq.~\ref{eq:3}), the input current $I_{\mathrm{in}}$ is integrated into the synaptic current $I_{\mathrm{syn}}$, where leakage decay governed by $\alpha$ occurs at every step. Next, the membrane potential $U$ retains a fraction $\beta$ of its prior value and integrates the updated $I_{\mathrm{syn}}$. If the result exceeds $U_{\mathrm{thr}}$, the neuron fires at the next step and resets; otherwise, $U$ continues to decay:
\begin{equation}
  \begin{aligned}
    I_{\mathrm{syn}}[t+1] &= \alpha I_{\mathrm{syn}}[t] + I_{\mathrm{in}}[t+1] \\
    U[t+1] &= \beta U[t] + I_{\mathrm{syn}}[t+1] - R
  \end{aligned}
  \label{eq:3}
\end{equation}

\section{Related Work}\label{sec:related_work}

Digital neuromorphic accelerators are broadly categorized into ASIC- and FPGA-based platforms, trading flexibility against performance and power efficiency.

Among ASIC-based SNN accelerators, Intel Loihi~\cite{ref37} enables large-scale, low-power processing through event-driven parallelism, while IBM TrueNorth \cite{ref19} targets inference via ultra-dense, low-leakage cores. ODIN~\cite{ref9} provides fully event-driven computation with configurable neuron dynamics covering the twenty Izhikevich behaviors, together with embedded online learning based on spike-driven synaptic plasticity (SDSP). MorphIC~\cite{ref11} adopts a quad-core, binary-weight architecture with online learning through stochastic spike-driven synaptic plasticity (S-SDSP), and ReckOn~\cite{ref10} implements a sub-mm$^{2}$ recurrent SNN architecture with on-chip learning driven by eligibility traces. NEXUS~\cite{ref20} scales this approach further by combining a 16-core architecture with a diamond interconnection topology. These trends indicate that, as SNNs move toward real-world workloads, multi-core support becomes an essential capability, enabling scalability through efficient inter-core communication as demonstrated by ChronoSpike~\cite{ref31}.

FPGA implementations, in turn, emphasize flexibility and low power consumption for edge and embedded applications. Spiker+~\cite{ref8} offers clock-driven computation with customizable neuron models and network topologies. More recently, ReSCom~\cite{ref32} introduced a reconfigurable FPGA-based SNN accelerator supporting IF, LIF, and synaptic neuron models within a unified architecture, further enabling runtime trade-offs among accuracy, latency, and energy consumption. Han et al.~\cite{ref21} employ a hybrid event-driven neuron and clock-driven synapse updating algorithm to balance performance and energy efficiency, whereas Li et al.~\cite{ref22} adopt an adaptive scheme that dynamically switches between clock-driven and event-driven computation to further improve power efficiency. SupraSNN~\cite{ref38} adopts a superscalar-inspired FPGA architecture with synapse-level parallelism and operation-based routing, jointly optimizing mapping and scheduling.

Existing SNN co-design frameworks typically target isolated layers of the design stack, with automated arithmetic precision exploration receiving comparatively little attention---an aspect that \emph{Flex-Plorer} explicitly addresses.

Specifically, SpikeExplorer~\cite{ref12} searches over network topologies and neuron models via multi-objective Bayesian optimization to balance accuracy, resource utilization, and power consumption; however, it keeps the arithmetic precision fixed throughout the exploration. SpikeFlow~\cite{ref34} operates at the microarchitecture level by co-designing a temporal-spatial systolic dataflow with compile-time \emph{Silence Cells} pruning to eliminate inactive neurons. While achieving high hardware efficiency, its operator precision remains predetermined rather than systematically explored. Addressing the communication layer, Xiao et al.~\cite{ref35} optimize inter-core spike routing on multi-core substrates using Hilbert-curve placement and runtime spike merging on the NoC, while treating the underlying compute datapath as a static black box.

\emph{Flex-Plorer} addresses the orthogonal dimension left largely unexplored by prior art:
automated, precision-level design-space exploration (DSE). Holding the target network
topology fixed, it searches over heterogeneous quantization parameters---feed-forward and
recurrent weight bit-widths, and leakage representation---using
a heuristic search method driven by a configurable, user-weighted cost metric. Evaluation
fidelity is preserved without costly floating-point approximation: accuracy is assessed by
a hardware- and dataflow-aware inference engine accelerated through
just-in-time compilation, while machine-learning-based estimators predict resource
utilization without per-candidate RTL synthesis. Finally, the
framework configures synthesizable RTL and driver software directly from the minimum-cost
configuration, closing the loop from algorithmic exploration to physical deployment.

\section{Flexi-NeurA}\label{sec:Flexi-NeurA}

\begin{figure}[htbp]
	\centering
	\includegraphics[width=\linewidth]{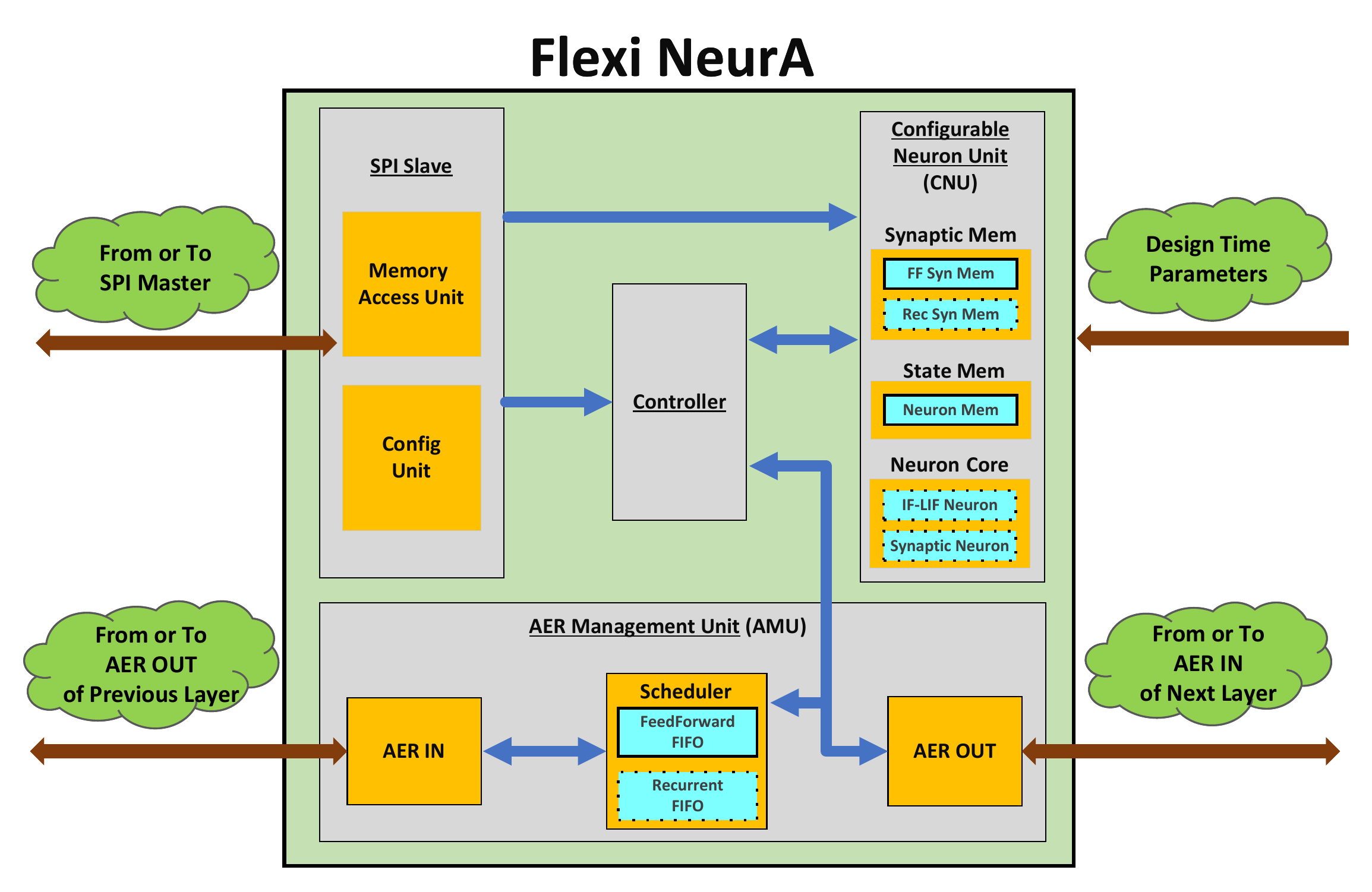}
	\caption{Top-level architecture of the Flexi-NeurA processing core and its main functional units.}
	\label{fig:core}
\end{figure}

Flexi-NeurA is built on four core architectural principles: time-multiplexed
computation, event-driven operation, multi-core organization, and design-time
configurability. These principles are jointly realized through a modular
processing core comprising an SPI slave, a controller, a Configurable Neuron
Unit (CNU), and an AER Management Unit (AMU) (Fig.~\ref{fig:core}). Because only
the components required by a given configuration are instantiated, this
modularity directly translates into reduced hardware overhead and lower power
consumption, while still supporting a wide range of neuron models and network
topologies.

Neuron updates in Flexi-NeurA follow a time-multiplexed execution model in which
a shared neuron-compute datapath sequentially processes all neurons in a layer.
For each neuron, the datapath integrates incoming spikes into the membrane
potential or the synaptic current, evaluates threshold crossing or leakage, and
then proceeds to the next neuron. This sequential reuse of compute resources
significantly reduces hardware overhead while preserving deterministic
execution.

Computation is event-driven and encoded via three address-based packet types.
ASPL ($9$~bits) carries a firing neuron's address to the next layer. ASCL
($8$~bits) conveys intra-layer spike addresses for recurrent connections at the
next time step. EOTS ($9$~bits) marks time-step completion; its $8$ address bits
identify the neuron with the highest accumulated membrane potential, while its
control bit signals step completion and---on the final step---triggers neuron
reset and leakage evaluation, obviating a separate End-of-Input packet.

Within each core, incoming ASPL/ASCL packets sequentially update neurons via
their synaptic weights; neurons crossing the threshold emit ASPL packets, while
the others undergo leakage. Embedding the highest-potential address in every
EOTS packet enables output identification by either spike count or accumulated
potential, with no extra packet type or side channel required.

\begin{figure}[htbp]
	\centering
	\includegraphics[width=\linewidth]{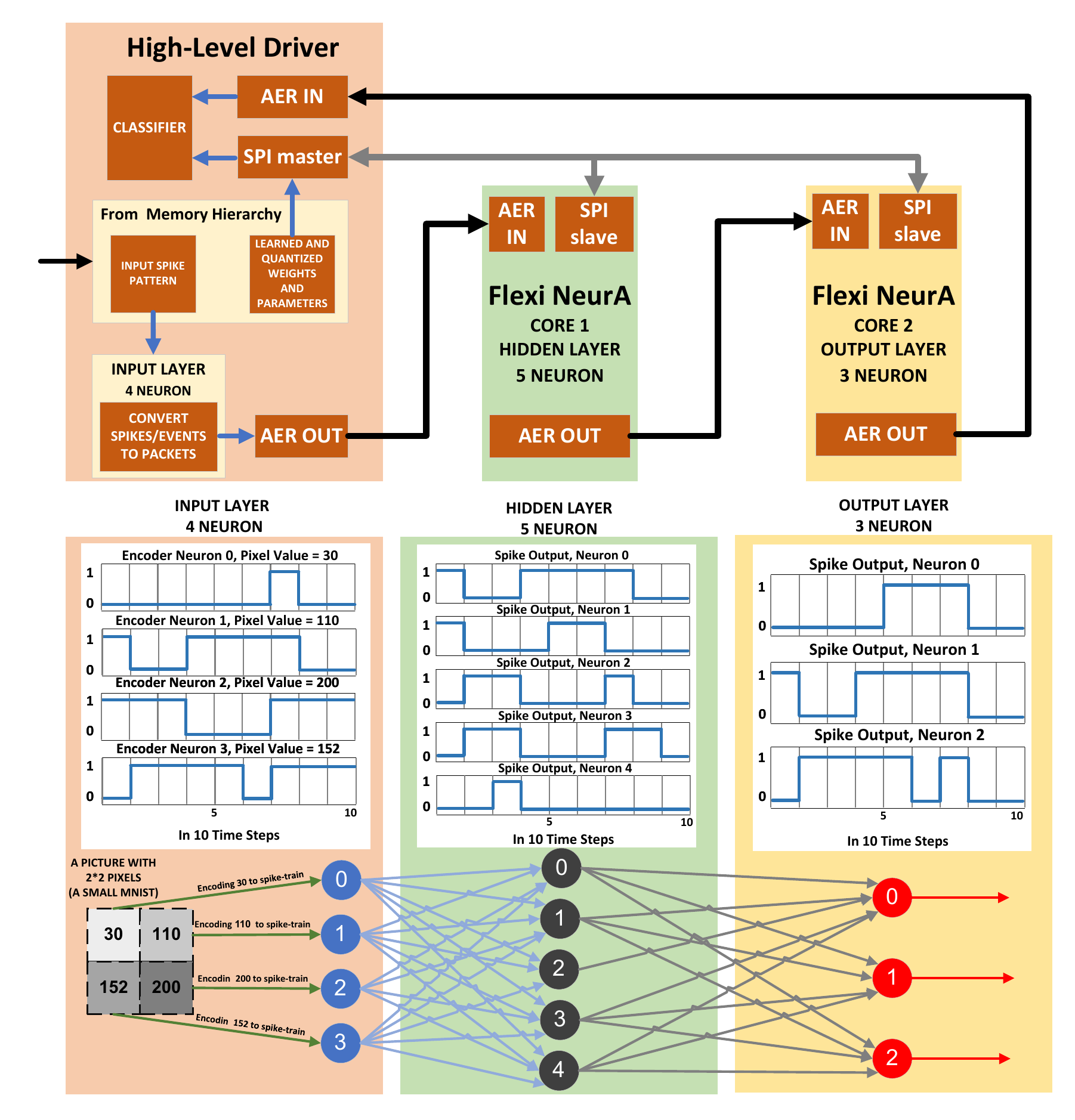}
	\caption{multi-core-compatible system organization showing AER-based inter-layer communication and layer-to-core mapping.}
	\label{fig:multicore}
\end{figure}

The network comprises an input layer, several hidden layers, and an output
layer, with each hidden and output layer mapped to a dedicated Flexi-NeurA core.
Inter-core communication relies exclusively on address-event packets---ASPL and
EOTS---transferred from the AER-OUT port of layer~$L$ to the AER-IN port of
layer~$L+1$, enabling layer-by-layer propagation without shared memories or side
channels, as shown in Fig.~\ref{fig:multicore}.

A high-level driver serves as the input layer, the SPI master, and the output
handler: it streams encoded spikes into the first core, issues EOTS packets per
time step, and reads back events from the final core for classification. The
predicted class can be identified via the spike count (by counting ASPL
packets), via the winning-neuron address embedded in EOTS (the highest membrane
potential at the last step), or via direct SPI read-out of the neuron
potentials. The same driver also configures each core over SPI, loading synaptic
weights and neuron parameters (model, leak, threshold, and active neuron count).

Flexi-NeurA exposes its core architectural choices, namely the neuron model, the
network topology, and the numerical precision, as HDL design-time parameters,
enabling application-specific specialization prior to synthesis. Because the
on-chip memory for synaptic weights, membrane potentials, and synaptic currents
is sized automatically as a function of the neuron count and the selected
bit-width, increasing the precision improves temporal accuracy and stability at
the cost of higher area and power, whereas reducing it lowers resource usage,
allowing designers to directly trade off accuracy, latency, and hardware cost
through a single set of parameters.

The supported network topologies include feedforward, recurrent All-to-All False,
and recurrent All-to-All True. Hardware support is provided for the Leaky
Integrate-and-Fire and Synaptic neuron models, while the Integrate-and-Fire
model is realized by disabling the leak term in the Leaky Integrate-and-Fire
implementation.

\subsection{Configurable Neuron Unit (CNU)}

The CNU is the core computational module in Flexi-NeurA and is responsible for
updating all neuron states. It is a highly modular unit composed of dedicated
memories for storing parameters and a computational core for executing the
neuron logic. The entire datapath of the CNU, including its memories and
computational units, is parameterized, meaning that its structure and
bit-precision are configured at design time to match the specific requirements
of the application, thereby optimizing hardware resource usage. As depicted in
Fig.~\ref{fig:cnu}, the CNU comprises three components---Synaptic Memory, Neuron
State Memory, and Neuron Core:

\begin{figure}[htbp]
	\centering
	\includegraphics[width=0.8\linewidth]{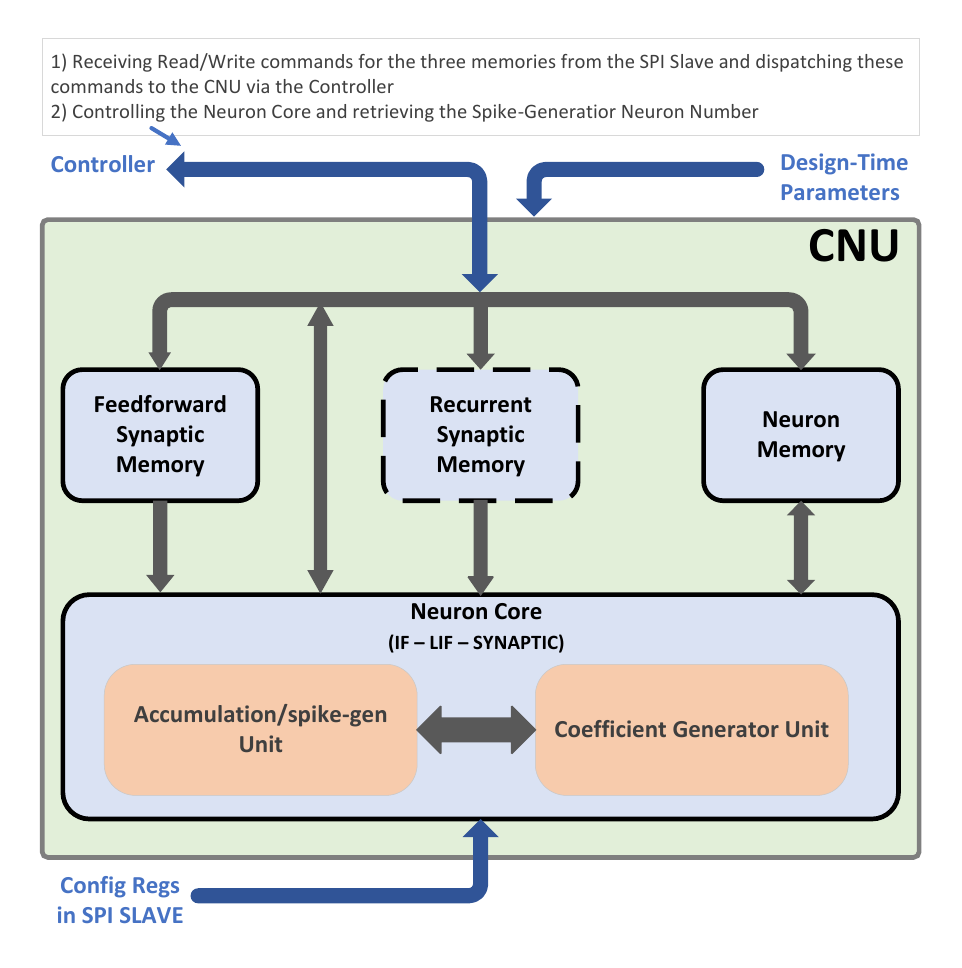}
	\caption{Block diagram of the Configurable Neuron Unit (CNU), illustrating the datapath and the interaction between the controller, the synaptic and neuron memories, and the Neuron Core.}
	\label{fig:cnu}
\end{figure}

\paragraph{Synaptic Memory}
This memory module comprises a feedforward Synaptic Memory for previous-layer
weights and, optionally, a Recurrent Synaptic Memory for same-layer weights. The
latter is instantiated only when the network is recurrent and adopts the ATA-T
topology, thereby avoiding unnecessary hardware overhead. Each memory is
organized into blocks of rows, with each row storing $8$~weights. Its dimensions
are automatically derived at design time from three parameters:
(i)~Number of Blocks, determined by the number of spike-source neurons
(presynaptic for feedforward, same-layer for recurrent) rounded up to the nearest
power of two to enable fixed-bit block addressing (e.g., $110$~neurons
$\rightarrow$ $128$~blocks, requiring a $7$-bit address);
(ii)~Rows per Block, defined by the number of destination neurons divided by $8$
and rounded up to the nearest power of two for row indexing (e.g., $90$~neurons
$\rightarrow$ $12$~required rows $\rightarrow$ $16$~allocated rows, requiring a
$4$-bit address); and
(iii)~Row Width, which is equal to $8$ times the user-defined weight
bit-precision (e.g., $4$-, $6$-, or $8$-bit weights).
This fully automated sizing ensures that the memory scales precisely according to
the network topology and the precision constraints, minimizing hardware
utilization while satisfying the performance requirements.

\paragraph{Neuron State Memory}
This RAM stores the dynamic state variables of each neuron in a dedicated row,
holding the membrane potential ($V_{\mathrm{m}}$) and the synaptic current
($I_{\mathrm{syn}}$)---and, where required, the potential accumulated over
multiple time steps---together. It is likewise auto-sized at design time along
two axes: (i)~Row Width, computed as the sum of the designer-specified
bit-widths of the state variables (e.g., $9$~bits for $V_{\mathrm{m}}$ and
$8$~bits for $I_{\mathrm{syn}}$, totaling $17$~bits), rounded up to the nearest
multiple of 8 bits ($24$~bits) for efficient access; and (ii)~Number of Rows, set by
the neuron count per layer, rounded up to the nearest power of two for
streamlined addressing. This design-time configurability ensures that the memory
footprint scales precisely with the layer size and the state precision.

\paragraph{Neuron Core}
The Neuron Core serves as the computational engine of the CNU, implementing the
selected neuron model (e.g., LIF, IF, or Synaptic) via design-time parametric
synthesis. It fetches parameters from the Synaptic Memory, the Neuron State
Memory, and the configuration registers, performs the required updates, and
writes the results back to the Neuron State Memory. The core executes two primary
operations. (i)~\textit{Integration}: for each incoming ASPL/ASCL packet, the
controller decodes the source address, fetches the corresponding quantized
weight, reads the destination neuron's state, and performs a single-neuron
update. Notably, only the targeted neuron's state is accessed and modified.
(ii)~\textit{Spike Generation and Leakage}: triggered by the control bit of the
last ASPL packet in a given time step, the neurons are processed sequentially in
a time-multiplexed manner. The state is read and compared against the threshold.
If the threshold is exceeded, a spike is registered and the configured reset is
applied. Otherwise, leakage is applied to the membrane potential
($V_{\mathrm{m}}$)---and to the synaptic current ($I_{\mathrm{syn}}$), if
applicable---via the Coefficient Generator Unit, before the updated state is
written back.

\begin{figure}[htbp]
	\centering
	\includegraphics[width=\linewidth]{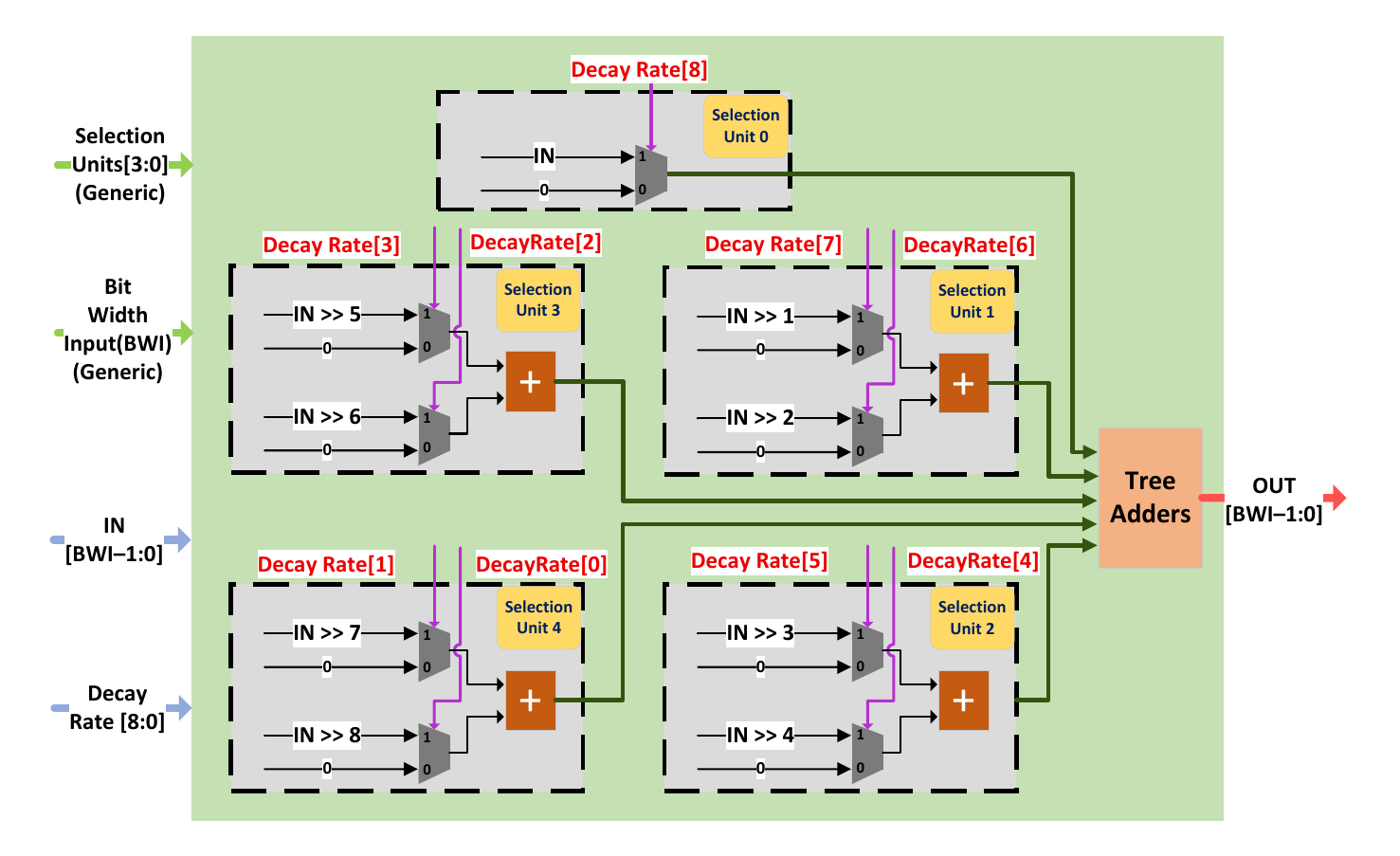}
	\caption{Coefficient Generator (CG) structure based on a shift-and-add network. The $9$-bit \texttt{DecayRate[8:0]} register configures selectable shifts to approximate any leakage factor in the $[0,1]$ range with $1/256$ resolution.}
	\label{fig:cg}
\end{figure}

\paragraph{Coefficient Generator (CG)}
This module models neuronal leakage without multipliers, enabling low-cost decay
in fixed-point hardware via a five-block shift-and-add network
(Fig.~\ref{fig:cg}): one bypass block and four data blocks with right-shift pairs
$(1,2)$, $(3,4)$, $(5,6)$, and $(7,8)$~bits, corresponding to Selection
Units~1--4. This structure approximates any decay factor $k/256 \in [0,1]$ (where
$k \in [0,255]$), achieving a resolution of $1/256$ with a worst-case rounding
error strictly below $1/512$.

The decay coefficient is encoded in a $9$-bit register, \texttt{DecayRate[8:0]}:
bit~$8$ activates the bypass path (Selection Unit~0), applying no decay for ideal
Integrate-and-Fire (IF) behavior, while bits~$7$--$0$ select the individual
shifts---bit~$7$ for $\texttt{IN} \gg 1$, bit~$6$ for $\texttt{IN} \gg 2$, down
to bit~$0$ for $\texttt{IN} \gg 8$---with each bit pair driving one adder block,
and the results subsequently summed via a tree adder. For example, a decay
factor of $0.59765625$ ($k = 153$) is encoded as \texttt{"010011001"}, activating
the $1/2$, $1/16$, $1/32$, and $1/256$ shift paths, whose sum yields the target
value.

Design-time configurability is provided through the $4$-bit
\texttt{SelectionUnits[3:0]} parameter, enabling the exclusion of unused blocks to
reduce area and power. Since the CG relies solely on shift-and-add operations, it
preserves signed-arithmetic symmetry even when blocks are omitted.

\subsection{Serial Peripheral Interface (SPI)}

The SPI module serves as the primary interface for an external master to
configure the core and access its internal memories, supporting two functions:
writing to the configuration registers and performing byte-level read/write
access to the three main memories---Neuron State, Feedforward Synaptic, and
Recurrent Synaptic.

\begin{figure}[htbp]
	\centering
	\includegraphics[width=\linewidth]{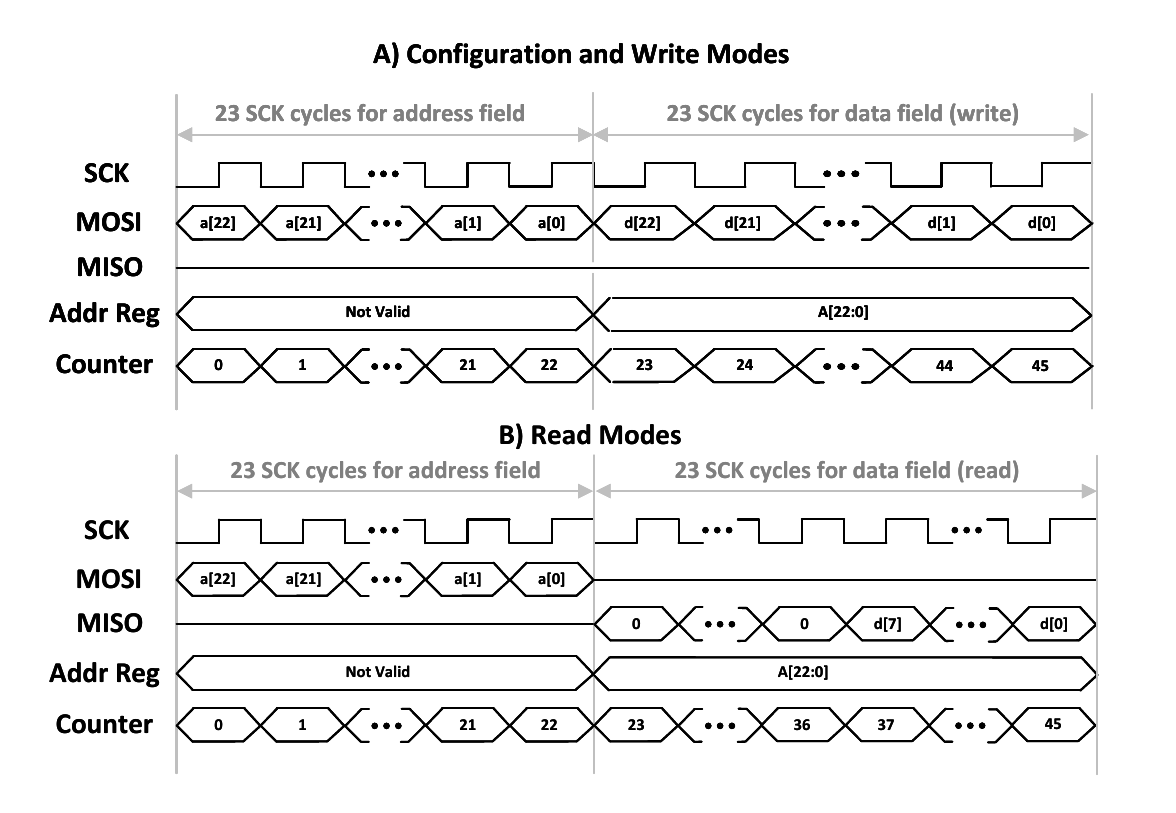}
	\caption{SPI communication timing. (A) Configuration and write operations using $23$-bit address and data fields. (B) Read operation with a $23$-bit address input and a $23$-bit data output via MISO.}
	\label{fig:spi}
\end{figure}

\paragraph{R/W Memory Operations}
All SPI transactions follow a fixed $46$-cycle frame comprising an address phase
(cycles $0$--$22$) and a data phase (cycles $23$--$45$), as shown in
Fig.~\ref{fig:spi}. During the address phase, a $23$-bit command is shifted in
via MOSI and latched into \texttt{AddrReg}. This word encodes: \texttt{bit[22]}
for the mode (memory access $=$ \texttt{'1'}, configuration $=$ \texttt{'0'});
\texttt{bit[21]} for the read/write direction; \texttt{bits[20:19]} for the target
memory (Neuron State $=$ \texttt{"00"}, Feedforward Synaptic $=$ \texttt{"01"},
Recurrent Synaptic $=$ \texttt{"10"}); and \texttt{bits[18:0]} for the payload,
containing either a register index or a memory address. During the data phase,
$23$~bits are received via MOSI for writes (Fig.~\ref{fig:spi}A), while $8$~bits
are transmitted via MISO for reads, beginning near cycle~$36$
(Fig.~\ref{fig:spi}B).

For the Neuron State Memory, the payload splits into \texttt{bits[7:0]} (neuron
row index) and \texttt{bits[18:8]} (byte offset within that neuron's state). For
the Synaptic Memory (feedforward or recurrent), the same command format applies
with a wider addressing map suited to larger arrays: \texttt{bits[12:0]} denote
the row index and \texttt{bits[18:13]} the byte offset. In both cases, $8$-bit
data is transferred during the second phase, yielding uniform byte-level access
across all three memories.

\paragraph{Configuration Registers}
Since a single master addresses multiple slave cores, each core is assigned a
unique design-time identifier fixed prior to synthesis. At the start of
configuration, the master broadcasts a value to the core-select register in all
cores simultaneously; each core compares this value against its own ID and, upon
a match, becomes the active target for the subsequent commands.

\begin{figure}[htbp]
	\centering
	\includegraphics[width=\linewidth]{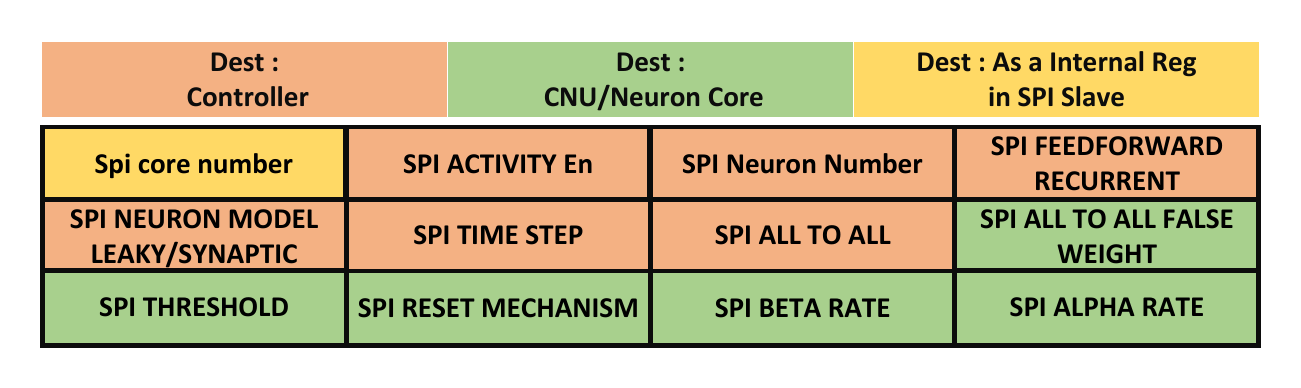}
	\caption{Configuration registers and their functional destinations within the Flexi-NeurA core.}
	\label{fig:config}
\end{figure}

While the design-time parameters fix the structural aspects of the accelerator
(neuron model, topology, and memory dimensions), the SPI module additionally
programs $12$ write-only, runtime-accessible configuration registers that govern
controller coordination, neuron-core control, and the internal SPI state
(Fig.~\ref{fig:config}).

\subsection{AER Management Unit (AMU)}

This unit manages the handshake-based reception and transmission of event packets
while preserving their processing order. It buffers packets arriving from the
previous layer alongside those generated locally in recurrent modes. The unit
comprises three submodules:

\paragraph{AER-IN}
Executes handshaking with the previous layer to receive ASPL and EOTS packets
and, upon ensuring available capacity in the feedforward scheduler, forwards them
to the scheduling queue.

\paragraph{AER-OUT}
Manages handshakes with the subsequent layer to transmit the locally generated
ASPL and EOTS packets.

\paragraph{Scheduler}
Comprises two FIFO-based queues. The \textit{feedforward queue} accepts external
ASPL and EOTS packets from AER-IN and forwards them to the controller in arrival
order. The \textit{recurrent queue} buffers locally generated ASCL packets for
reuse in recurrent computations; notably, this queue is instantiated exclusively
for ATA-T topologies.

\subsection{Controller}

The controller is the core's central operational unit, managing all processes
through four main phases---(a) configuration, (b) spike integration, (c) leakage
and spike generation, and (d) packet transmission:

\paragraph{Configuration Phase}
Assigned the highest priority, this phase initializes the core via SPI. A
controller state machine decodes the incoming opcode to select the target memory
path and enters one of six states (e.g., \texttt{R/W\_NEUR},
\texttt{R/W\_FF\_SYN}, \texttt{R/W\_REC\_SYN}) to perform a single-cycle
read/write access to the feedforward synaptic, recurrent synaptic, or neuron
state memory, ensuring deterministic loading of weights and neuron parameters.
After each access, the controller transitions to a \texttt{WAIT\_SPI} state to
synchronize with the slower SPI bus, guaranteeing transfer completion and
preventing timing-related data corruption.

\paragraph{Spike Integration Phase}
During this phase, the controller processes the incoming spike event packets
(ASPL and ASCL) queued in the schedulers. This phase is divided into two main
sub-operations:
(i)~Feedforward integration: the controller services ASPL packets dequeued from
the feedforward FIFO. For each packet, it enters the \texttt{FF\_INTEG} state,
decodes the source address, and---using a time-multiplexed compute
unit---sweeps the current layer neuron by neuron. For each target neuron, it
fetches the corresponding synaptic weight and current state
($V_{\mathrm{m}}$/$I_{\mathrm{syn}}$), accumulates the contribution, and writes
back the updated state. After all neurons are updated, it advances to the next
ASPL packet.
(ii)~Recurrent integration: upon receiving an EOTS token, the controller enters
the \texttt{REC\_INTEG} state to service ASCL packets from the recurrent
scheduler based on the configured topology: \texttt{ATA-T} (All-to-All True) or
\texttt{ATA-F} (All-to-All False). In the \texttt{ATA-T} mode, for each ASCL
packet, the time-multiplexed compute unit sweeps all neurons, reading the
recurrent weight and the current neuron state, accumulating the contribution to
$V_{\mathrm{m}}$/$I_{\mathrm{syn}}$, and writing it back to the Neuron State
Memory. In the \texttt{ATA-F} mode, only the source neuron is updated using its
self-recurrent weight. This processing proceeds packet by packet and neuron by
neuron until the ASCL queue is completely drained.

\begin{figure}[htbp]
	\centering
	\includegraphics[width=\linewidth]{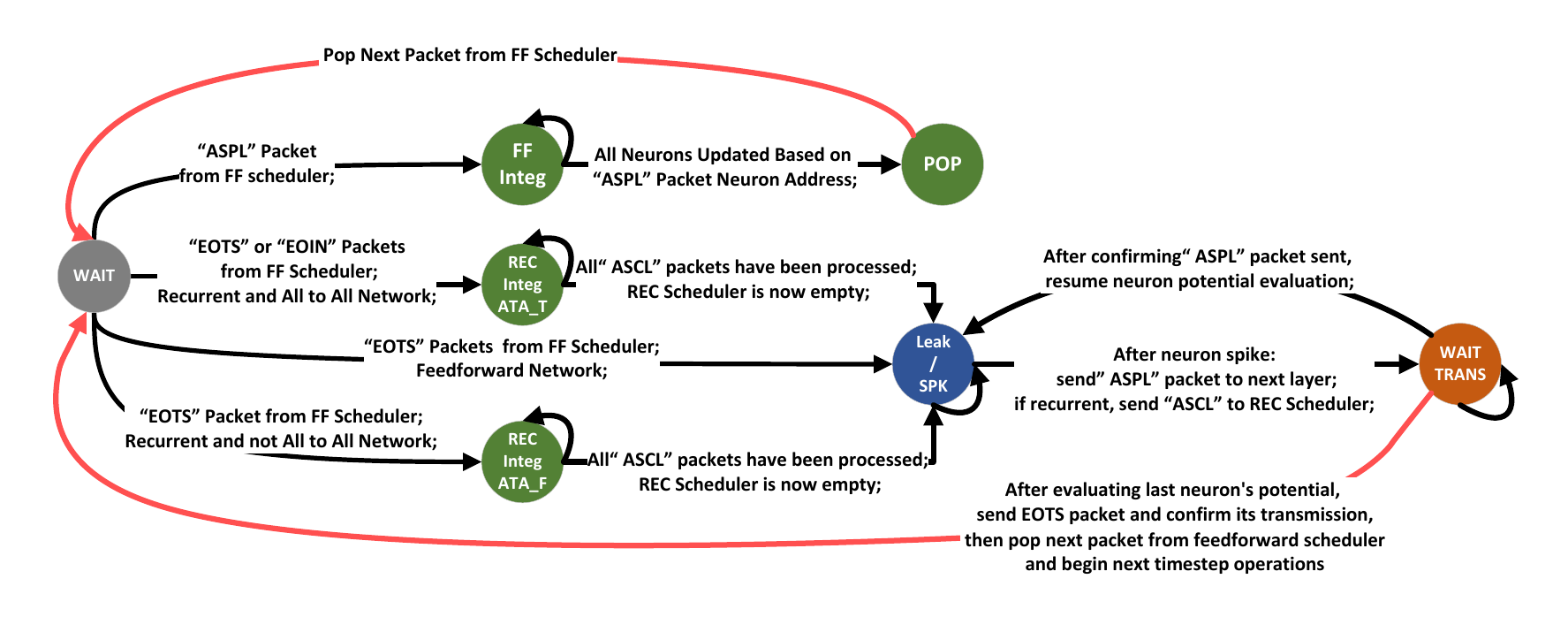}
	\caption{State machine for the controller's spike integration and leakage and spike generation phases. Incoming spike packets from the feedforward or recurrent paths trigger per-neuron updates, followed by potential evaluation, spike emission, and control signaling.}
	\label{fig:controller2}
\end{figure}

\paragraph{Leakage and Spike Generation Phase}
After all ASPL and ASCL effects are integrated for a given time step, the
controller transitions to the \texttt{LEAK\_SPK} state
(Fig.~\ref{fig:controller2}) and sequentially evaluates the neurons using the
time-multiplexed unit. For each neuron, if its membrane potential
($V_{\mathrm{m}}$) reaches or exceeds the threshold ($V_{\mathrm{th}}$), a spike
is generated. This triggers the transmission of an ASPL packet to the subsequent
layer (and locally enqueues an ASCL packet if the network is recurrent), followed
by a membrane reset. Otherwise, leakage is applied to $V_{\mathrm{m}}$ (and to
the synaptic current, $I_{\mathrm{syn}}$, if modeled). At the final time step of
a sample, signaled by the control bit of the \texttt{EOTS} packet, a
\textit{lazy reset} mechanism writes zeros directly into the neuron's state in
place of the computed values, thereby avoiding costly global clear operations and
preparing the core for the next input sample.

\paragraph{Packet Transmission Phase}
Following spike generation, the controller enters the \texttt{WAIT\_TRANS} state
(Fig.~\ref{fig:controller2}) to synchronize with the subsequent layer. To
naturally handle backpressure and ensure lossless data transfer, it stalls the
generated packet while the feedforward queue of the next layer is full,
transmitting it only when space becomes available. Once all spikes for the
current evaluation cycle are processed, the controller issues an \texttt{EOTS}
(End of Time Step) packet to the next layer. Its control bit is cleared if
further time steps remain, or asserted if the current step concludes the
sample---thereby signaling the subsequent layer to perform its own final
processing and lazy reset.

\begin{figure}[htbp]
	\centering
	\includegraphics[width=0.5\linewidth]{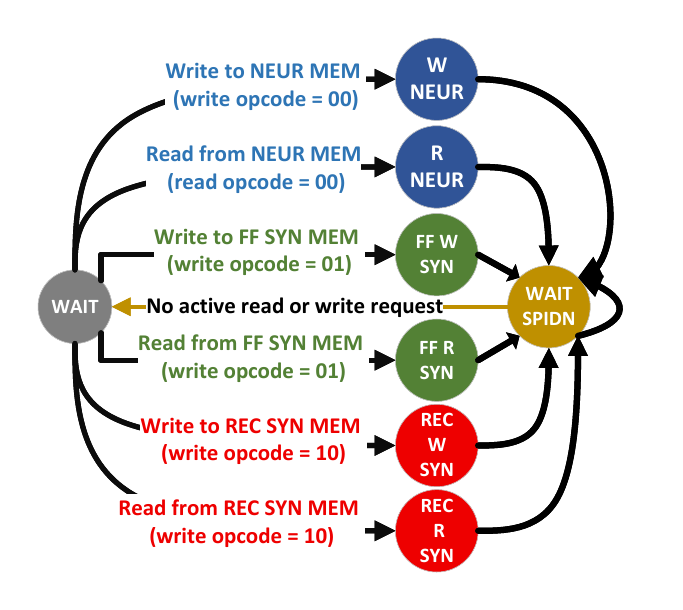}
	\caption{Controller state machine for handling read/write operations across the neuron, feedforward synapse, and recurrent synapse memories, including SPI wait synchronization.}
	\label{fig:controller1}
\end{figure}

\section{Flex-Plorer}\label{sec:Flex-Plorer}

Flexi-NeurA provides configurable network topology, neuron model,
weight bit-width, leakage-parameter precision, core count, and
neuron count per core. Together, these options define a large
hardware design space. Flex-Plorer (Fig.~\ref{fig:Flex-Plorer})
explores this space through a hardware--software co-design
framework that links a software-level SNN description to the
Flexi-NeurA architecture.

The workflow has four interconnected units: the graphical user
interface (GUI), the SNN training unit, the design-space
exploration (DSE) unit, and the RTL configurator. The GUI collects
the network architecture, neuron model, target workload, and
design constraints. It also defines the search ranges for
feedforward and recurrent weight bit-widths and the two decay
precisions. The training unit trains the network once and passes
the trained weights and decay-parameter values to DSE.

DSE combines the trained parameters with the architecture and
search-space settings from the GUI. It evaluates each candidate
using a hardware-aware accuracy simulator and hardware-resource
estimators. A heuristic search seeks a feasible configuration
that minimizes the specified cost while meeting the user's
constraints. The RTL configurator then applies the selected
parameters to the Flexi-NeurA HDL templates and generates the
configuration-specific HDL and deployment artifacts.

\subsection{Graphical User Interface (GUI)}

The GUI lets the user define the SNN application and hardware
design space. Its inputs include the target workload and dataset,
network topology, neuron model, training settings, and search
ranges for the hardware parameters. These parameters include
weight bit-width, leakage-parameter precision, the number of
processing cores, and the number of neurons per core.

The user sets the relative importance of inference accuracy and
hardware cost through $C_A$ and $C_H$, respectively, with
$C_A+C_H=1$. The GUI passes the network and training settings to
the training unit, and the design-space parameters, constraints,
and cost-function coefficients to DSE.

\subsection{SNN Training and Model Preparation}

The training unit uses snnTorch~\cite{ref14} to train the selected
network once with backpropagation through time (BPTT). It supports
the neuron models and network topologies compatible with
Flexi-NeurA. After training, the synaptic weights and neuron
leakage parameters are extracted in floating-point format.
These values are passed to DSE for evaluation under the candidate
fixed-point configurations defined through the GUI.

\subsection{Design-Space Exploration}

A Flexi-NeurA configuration specifies the network topology,
network dimensions, neuron model, and numerical precisions.
We represent it as
\begin{equation}
  c =
  \left\{
  \underbrace{
  T, L, \mathbf{N}, P_s, M, K
  }_{\text{GUI-fixed parameters}},
  \underbrace{
  P_{w}^{\mathrm{ff}}, P_{w}^{\mathrm{rec}},
  P_{\alpha}, P_{\beta}
  }_{\text{DSE parameters}}
  \right\},
  \label{eq:configuration}
\end{equation}
where $T$ is the network topology, $L$ is the number of layers,
and $\mathbf{N}=(N_1,\ldots,N_L)$ gives the neuron count in each
layer. The parameters $M$ and $K$ specify the neuron model and
the number of processing cores, respectively. Neuron-state
precision is denoted by $P_s$. The feedforward and recurrent
weight precisions are $P_{w}^{\mathrm{ff}}$ and
$P_{w}^{\mathrm{rec}}$, while $P_{\alpha}$ and $P_{\beta}$
specify the precisions of the leakage parameters $\alpha$
and $\beta$.

To limit the search space, the user fixes the architecture
parameters $(T,L,\mathbf{N},M,K)$, neuron-state precision $P_s$,
and all other non-swept settings through the GUI. Independent
search ranges are then defined for four variables:
$P_{w}^{\mathrm{ff}}$, $P_{w}^{\mathrm{rec}}$, $P_{\alpha}$,
and $P_{\beta}$. Only these precision variables change during
DSE; the architecture and remaining settings stay fixed.

Two tools evaluate each candidate. The hardware-aware fixed-point
simulator measures inference accuracy on the target workload
using Flexi-NeurA's arithmetic and neuron-update order. The
hardware estimator predicts LUT and FF utilization with
synthesis-trained machine-learning models and BRAM utilization
with an analytical model. This allows candidates to be evaluated
without a synthesis run for each one.

DSE minimizes a weighted cost with two terms.
$\mathit{AccCost}$ penalizes the shortfall of
$\mathit{Hardware\_Aware\_Accuracy}$ from one, and
$\mathit{HwCost}$ combines the normalized LUT, FF, and BRAM
estimates. The user-defined coefficients $C_{\mathrm{A}}$
and $C_{\mathrm{H}}$ set the balance between accuracy and
hardware cost. Within the hardware term, $C_{\mathrm{LUT}}$,
$C_{\mathrm{FF}}$, and $C_{\mathrm{BRAM}}$ set the relative
importance of each resource:
\begin{align}
  \mathit{TotalCost} &= \mathit{HwCost} + \mathit{AccCost} \label{eq:total_cost} \\
  \mathit{AccCost}   &= C_{\mathrm{A}} \times \left(1 - \mathrm{Hardware\_Aware\_Accuracy}\right) \label{eq:acc_cost} \\
  \mathit{HwCost}    &= C_{\mathrm{H}} \times \left(C_{\mathrm{LUT}}\,\mathit{LUT}_{n} + C_{\mathrm{FF}}\,\mathit{FF}_{n} + C_{\mathrm{BRAM}}\,\mathit{BRAM}_{n}\right) \label{eq:hw_cost}
\end{align}
The weights satisfy the normalization constraints
\begin{equation}
  \begin{cases}
    C_{\mathrm{H}} + C_{\mathrm{A}} = 1 \\
    C_{\mathrm{LUT}} + C_{\mathrm{FF}} + C_{\mathrm{BRAM}} = 1
  \end{cases}
  \label{eq:constraints}
\end{equation}

Here, $\mathit{Hardware\_Aware\_Accuracy}$ is the accuracy
obtained from the fixed-point simulator. The quantities
$\mathit{LUT}_{n}$, $\mathit{FF}_{n}$, and $\mathit{BRAM}_{n}$
are the normalized resource estimates. The resulting total cost
expresses the chosen accuracy--hardware trade-off, with lower
values preferred.

At each search step, a heuristic method selects a candidate,
evaluates its accuracy and hardware cost, and uses the result
to guide the next selection. This process continues until the
minimum-cost configuration is found. The following subsections
describe the simulator, hardware estimation, and search procedure.

\subsubsection{Hardware-Aware Simulation}

The simulator models Flexi-NeurA's fixed-point arithmetic and
its time-multiplexed execution, in which neurons are processed
one at a time. It therefore accounts for both hardware arithmetic
and dataflow. The quantized weight bit-width directly affects
hardware execution time. Numba-based just-in-time compilation
[23] speeds up inference and reduces simulation overhead
during DSE.

\subsubsection{Hardware Estimation}
\label{sec:hardware_estimation}

The hardware-estimation module predicts the resources required
by each candidate SNN configuration without a separate synthesis
run. It uses machine-learning models trained on synthesis
measurements to estimate LUT and flip-flop utilization.
BRAM utilization is estimated separately with an analytical
memory model.

\paragraph{LUT and Flip-Flop Estimation}
\label{sec:lut_ff_estimation}
\mbox{}\par\nobreak

\begin{figure*}[p]
  \centering
  \includegraphics[
    width=\textwidth,
    height=0.82\textheight,
    keepaspectratio
  ]{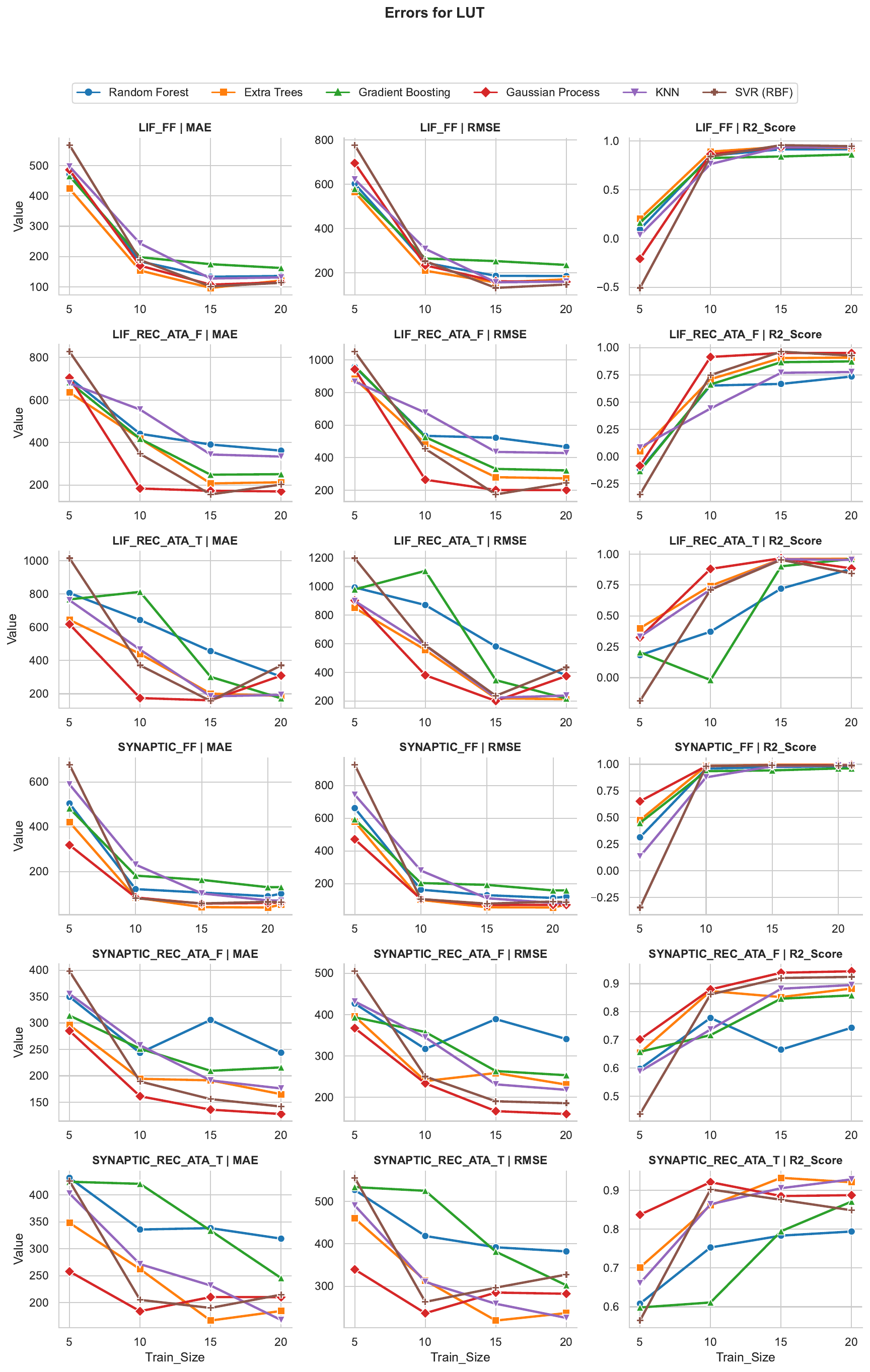}
  \caption{LUT estimation versus training-set size for Random
  Forest (RF), Extra Trees (ET), Gradient Boosting (GB),
  Gaussian Process (GP), $k$-Nearest Neighbors (KNN), and
  RBF-kernel Support Vector Regression (SVR).
  Columns report MAE, RMSE, and $R^2$, from left to right.
  Rows show LIF Feed-Forward, LIF ATA-F, LIF ATA-T,
  SYNAPTIC Feed-Forward, SYNAPTIC ATA-F, and SYNAPTIC ATA-T,
  from top to bottom. The horizontal axis gives the
  training-set size (5, 10, 15, and 20 samples).}
  \label{fig:lut_evaluation}
\end{figure*}

\begin{figure*}[p]
  \centering
  \includegraphics[
    width=\textwidth,
    height=0.82\textheight,
    keepaspectratio
  ]{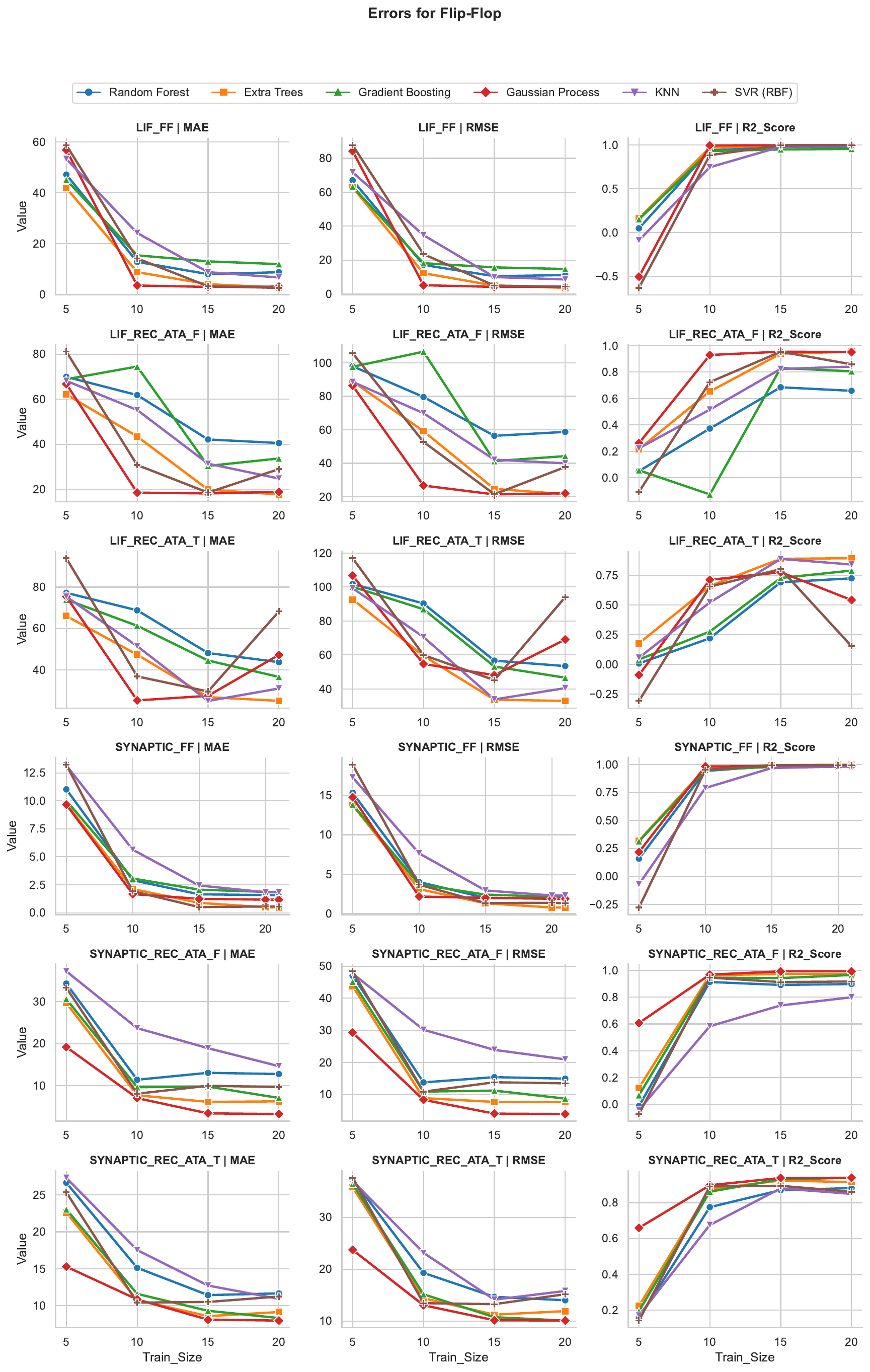}
  \caption{Flip-flop estimation versus training-set size for
  RF, ET, GB, GP, KNN, and SVR, using the same panel
  arrangement as Fig.~\ref{fig:lut_evaluation}.
  Columns report MAE, RMSE, and $R^2$, from left to right.
  Rows show LIF Feed-Forward, LIF ATA-F, LIF ATA-T,
  SYNAPTIC Feed-Forward, SYNAPTIC ATA-F, and SYNAPTIC ATA-T,
  from top to bottom. The horizontal axis gives the
  training-set size (5, 10, 15, and 20 samples).}
  \label{fig:ff_evaluation}
\end{figure*}

\begin{adjustwidth}{1.5em}{0pt}

\estimationheading{Configurations and Dataset}

The LUT and flip-flop estimators are trained on resource
measurements from designs synthesized with different feedforward
and recurrent weight precisions.

The dataset covers six neuron--topology configurations: three
topologies---Feed-Forward, Recurrent (ATA-F), and Recurrent
(ATA-T)---combined with two neuron-model categories, LIF/IF
and SYNAPTIC. IF and LIF share one category because their
synthesis results are identical in the implementations considered
here. SYNAPTIC is treated separately.

For each configuration, approximately 30 samples are obtained
by synthesizing a single Flexi-NeurA core. The neuron counts
in the previous and current layers remain fixed, while
feedforward and, where applicable, recurrent weight precisions
vary from 4 to 64 bits. Each sample records these precisions
and the resulting LUT, flip-flop, and BRAM utilization.
The LUT and flip-flop measurements are used as regression
targets; BRAM utilization is estimated separately with an
analytical model.

Changes in leakage precision affect the resources used by the
compact shift-and-add CG. The estimator treats this as a
secondary effect and neglects it in the cost model, so
$P_{\alpha}$ and $P_{\beta}$ are not included as inputs.
Their effects on quantized neuron dynamics and inference
accuracy are still modeled by the hardware-aware simulator.

For Feed-Forward configurations, the input is the feedforward
weight precision:
\[
\mathbf{x}_i =
\left[
P_{w,i}^{\mathrm{ff}}
\right].
\]
For recurrent configurations, the input includes both weight
precisions:
\[
\mathbf{x}_i =
\left[
P_{w,i}^{\mathrm{ff}},
P_{w,i}^{\mathrm{rec}}
\right].
\]

A separate predictor is trained for each resource type and
neuron--topology configuration:
\[
\widehat{y}^{(r,q)}
=
f_{r,q}(\mathbf{x}),
\]
where $r$ identifies the resource, either LUT or flip-flop,
and $q$ identifies one of the six configurations. During DSE,
each predictor uses the candidate's weight precisions to
estimate resource utilization without another synthesis run.

\estimationheading{LUT and Flip-Flop Estimation Models}

We evaluate six regression models: Random Forest (RF),
Extra Trees (ET), Gradient Boosting (GB), Gaussian Process
(GP) regression, $k$-Nearest Neighbors (KNN), and Support
Vector Regression (SVR) with an RBF kernel. Each predicts
LUT or flip-flop utilization from the precision inputs
defined above.

RF averages decision trees trained on bootstrap samples,
capturing non-linear and piecewise relationships between
precision and resource usage. ET also averages decision
trees, but selects split thresholds randomly rather than
optimizing them. This additional randomization reduces
prediction variance. GB builds shallow trees sequentially,
fitting each to the residual errors of the current prediction
and combining them in a weighted sum.

GP is a nonparametric probabilistic model that conditions
a prior over functions on the training data. Its kernel
defines similarity between configurations and the smoothness
of the fitted trend. KNN instead makes local, distance-based
predictions, typically by averaging measurements from the
nearest training configurations without fitting an explicit
functional form. RBF-kernel SVR fits a function within a
margin of the training targets, representing non-linear
relationships without explicit polynomial features.

\estimationheading{LUT and Flip-Flop Estimation Results}

For each of the six neuron--topology configurations, a single
core is synthesized at 30 different bit-precision settings.
The resulting samples are split into 20 for training and
10 for evaluation. The estimators are trained incrementally
on 5, 10, 15, and 20 samples from the training set. At each
step, predictions are compared with the same 10 evaluation
samples.

Figures~\ref{fig:lut_evaluation} and~\ref{fig:ff_evaluation}
show the LUT and flip-flop results for RF, ET, GB, GP, KNN,
and SVR. In both figures, the columns report MAE, RMSE,
and $R^2$, from left to right. The rows cover FF, ATA-F,
and ATA-T for LIF, followed by the same topologies for
SYNAPTIC. The horizontal axis gives the number of training
samples.

Feed-forward (FF) configurations can be estimated accurately
with relatively few samples. For LUT utilization, $R^2$
ranges from about 0.75 to 0.99 with 10 training samples;
with 20 samples, even the lowest value is about 0.86.
For flip-flops, all models reach $R^2 \geq 0.94$ with
20 samples. Performance is less consistent with only five
samples: for LIF FF, GP and SVR yield negative $R^2$ values
of about $-0.2$ and $-0.5$ for LUTs, and $-0.5$ and $-0.65$
for flip-flops, respectively.

LUT estimation for recurrent networks depends more strongly
on the configuration. For LIF ATA-F, GP is the only model
to exceed $R^2=0.9$ with 10 samples, reaching about 0.91,
while RF remains near 0.65. With 20 samples, $R^2$ ranges
from about 0.74 (RF) to 0.94 (GP) for ATA-F in both neuron
types. The corresponding ranges are 0.84 (SVR) to 0.96 (ET)
for LIF ATA-T and 0.79 (RF) to 0.93 (KNN) for SYNAPTIC ATA-T.
More training samples do not always improve the estimates.
For example, SVR's $R^2$ for LIF ATA-T LUT utilization
reaches about 0.95 with 15 samples but falls to about 0.84
with 20.

Recurrent flip-flop estimation errors are more variable
at 20 samples. For LIF ATA-F, GB's $R^2$ falls to about
$-0.1$ with 10 samples before recovering to 0.81 with 20.
For LIF ATA-T, ET gives the lowest MAE at 20 samples
(about 25). In contrast, SVR's MAE increases from about
29 with 15 samples to about 68 with 20, while its $R^2$
drops from about 0.8 to 0.15. For SYNAPTIC ATA-F at
20 samples, GP has the lowest MAE (about 3.5) and KNN
the highest (about 15).

GP, the only Bayesian model evaluated here, does not
consistently outperform the other estimators. It achieves
the highest $R^2$ for SYNAPTIC FF LUT estimation with
five samples: about 0.65, compared with at most 0.48 for
the other models. It also gives the highest $R^2$ for
SYNAPTIC ATA-F LUT estimation with 15 to 20 samples
(about 0.94), with an MAE of about 125 at 20 samples.
For LIF ATA-F and SYNAPTIC ATA-F, its flip-flop errors
are the lowest or tied for the lowest with 10 to 20
samples. However, GP performs worst for LIF FF with
five samples. For LIF ATA-T with 20 samples, its LUT
MAE is about 310 ($R^2 \approx 0.88$), while its
flip-flop $R^2$ falls from about 0.78 to 0.54.
These comparisons use point estimates only; GP's
predictive uncertainty is not evaluated here.

\end{adjustwidth}

\paragraph{BRAM Estimation}
\mbox{}\par\nobreak

\begin{adjustwidth}{1.5em}{0pt}

\estimationheading{BRAM Utilization Estimation}

BRAM utilization is estimated directly from the required memory
dimensions and supported FPGA memory primitives, rather than
through regression. The estimator constructs each memory domain
and uses dynamic programming to allocate RAMB36 and RAMB18 blocks.
It reports utilization in 36K-equivalent blocks:
\begin{equation*}
B_{\mathrm{eq36k}}
=
N_{\mathrm{RAMB36}}
+
0.5N_{\mathrm{RAMB18}}.
\end{equation*}

Three memory domains contribute to the estimate:
(i)~neuron-state memory,
(ii)~feedforward synaptic memory, and
(iii)~recurrent-weight memory.
For each core, the total is
\begin{equation*}
B_{\mathrm{core}}
=
B_{\mathrm{neuron}}
+
B_{\mathrm{ff\text{-}syn}}
+
B_{\mathrm{rec\text{-}w}}.
\end{equation*}
All supported configurations include neuron-state and feedforward
synaptic memory. Only ATA-T uses a separate recurrent-weight
domain; this contribution is zero for Feed-Forward and ATA-F
mappings. This difference reflects the memory organization,
not an absence of recurrent connectivity in ATA-F. Summing
the contributions of all cores gives the network estimate.

The BRAM count depends on numerical bit-widths, neuron model,
topology, and the neuron counts in adjacent layers. Each layer
is computed by one core, so the previous- and current-layer
neuron counts determine the depths of that core's memory domains
and strongly affect its BRAM requirements. Denoting these counts
by $N_{\mathrm{prev}}$ and $N_{\mathrm{curr}}$, their
power-of-two padded dimensions are
\begin{equation*}
N_{\mathrm{prev}}'
=
2^{\lceil \log_2 N_{\mathrm{prev}}\rceil},
\qquad
N_{\mathrm{curr}}'
=
2^{\lceil \log_2 N_{\mathrm{curr}}\rceil}.
\end{equation*}
Each memory has a depth $D$, measured in addressable locations,
and a width $W$, measured in bits per location. The bit-widths
of the membrane potential, synaptic current, feedforward weights,
and recurrent weights are denoted by $P_v$, $P_i$,
$P_{\mathrm{ff}}$, and $P_{\mathrm{rec}}$, respectively.

Neuron-state memory holds the states of the current core's
neurons. For both IF/LIF and SYNAPTIC neurons, its depth is
therefore $N_{\mathrm{curr}}'$. The logical state width is
$P_v$ for IF/LIF and $P_v+P_i$ for SYNAPTIC. The stored word
width is rounded up to a multiple of eight:
\begin{align*}
(D_{\mathrm{neuron}},W_{\mathrm{neuron}})_{\mathrm{IF/LIF}}
&=
\left(
N_{\mathrm{curr}}',
8\left\lceil\frac{P_v}{8}\right\rceil
\right), \\
(D_{\mathrm{neuron}},W_{\mathrm{neuron}})_{\mathrm{SYNAPTIC}}
&=
\left(
N_{\mathrm{curr}}',
8\left\lceil\frac{P_v+P_i}{8}\right\rceil
\right).
\end{align*}
For SYNAPTIC neurons, the membrane potential and synaptic
current thus share a byte-aligned state word. Both neuron
categories use power-of-two depth padding.

The dimensions of the feedforward synaptic memory and the
ATA-T recurrent-weight memory are
\begin{align*}
D_{\mathrm{ff\text{-}syn}}
&=
\left\lceil\frac{N_{\mathrm{curr}}'}{8}\right\rceil
N_{\mathrm{prev}}',
&
W_{\mathrm{ff\text{-}syn}}
&=8P_{\mathrm{ff}}, \\
D_{\mathrm{rec\text{-}w}}
&=
\left\lceil\frac{N_{\mathrm{curr}}'}{8}\right\rceil
N_{\mathrm{curr}}',
&
W_{\mathrm{rec\text{-}w}}
&=8P_{\mathrm{rec}}.
\end{align*}
Each row stores weights for eight destination neurons.
A source neuron therefore requires
$\lceil N_{\mathrm{curr}}'/8\rceil$ rows. Multiplying this
row count by the padded source-neuron count gives the depth:
$N_{\mathrm{prev}}'$ for feedforward weights and
$N_{\mathrm{curr}}'$ for recurrent weights.

For each domain, the estimator considers supported RAMB36
and RAMB18 configurations that meet the required True Dual-Port
(TDP) or Simple Dual-Port (SDP) access mode. Each primitive
option $p$ has depth $d_p$, width $w_p$, and equivalent cost
$c_p$. To cover a requested depth $D$, the required number
of cascaded blocks and their combined cost are
\begin{equation*}
k_p=\left\lceil\frac{D}{d_p}\right\rceil,
\qquad
C_p(D)=k_pc_p,
\end{equation*}
where $c_p=1$ for RAMB36 and $c_p=0.5$ for RAMB18.
Each resulting depth-cascaded segment provides $w_p$ bits
of width.

Dynamic programming selects the minimum-cost combination
of segments that covers the requested width $W$:
\begin{align*}
\mathrm{DP}[0] &= 0, \\
\mathrm{DP}[w]
&=
\min_{p\in\mathcal{P}}
\left\{
\mathrm{DP}[\max(0,w-w_p)]+C_p(D)
\right\},
\qquad 1\leq w\leq W,
\end{align*}
where $\mathcal{P}$ is the set of enabled primitive options
for the selected access mode. The estimated utilization
of the domain is $\mathrm{DP}[W]$. Each state also tracks
the RAMB36 and RAMB18 counts. When allocations have the
same equivalent cost, the one with fewer physical blocks
is selected. Domains are packed independently; unused
capacity in one domain is not assumed to be available
to another.

The estimator offers automatic mode selection,
\texttt{force\_tdp}, and \texttt{force\_sdp}.
In automatic mode, it chooses the lowest-cost allocation
among access modes compatible with the domain:
\begin{align*}
B_{\mathrm{selected}}
&=
\min_{m\in\mathcal{M}_{\mathrm{compatible}}} B_m, \\
\mathcal{M}_{\mathrm{compatible}}
&\subseteq \{\mathrm{TDP},\mathrm{SDP}\}.
\end{align*}
An allocation is excluded if its port behavior does not
meet the architecture's requirements, even if it costs less.

Additional options enable or disable RAMB18 blocks, round
physical packing widths up to multiples of nine, or expand
64-bit SDP requests to 72 bits for an ECC-compatible layout.
These packing options must match the target implementation.
They do not change the architecture's byte-aligned state-word
definition, and a 72-bit allocation alone does not mean that
ECC is enabled.

The estimator finds the minimum allocation for the specified
memory organization and primitive choices; it does not
guarantee the mapping that synthesis will produce. The
estimate therefore applies only when its depths, widths,
banking, and access modes match the implementation.
Because memory is allocated in discrete primitives, BRAM
utilization changes in steps rather than continuously.

\estimationheading{BRAM Evaluation}

Figure~\ref{fig:bram_topology_evaluation} compares the analytical
BRAM estimates with reference values for six neuron--topology
configurations. The layer dimensions are fixed at PRE\,=\,256
and CURR\,=\,128, where PRE and CURR denote the previous-
and current-layer neuron counts, respectively.

ATA-F gives the lowest errors in both neuron categories:
MAE/RMSE are approximately $0.15/0.27$ for LIF and
$0.07/0.18$ for SYNAPTIC. LIF Feed-Forward has the highest
MAE ($\approx 1.23$), whereas LIF ATA-T has the highest
RMSE ($\approx 2.44$). For both Feed-Forward and ATA-T,
errors are lower with SYNAPTIC than with LIF. The
Feed-Forward MAE/RMSE values are approximately $0.37/0.71$
for SYNAPTIC, compared with $1.23/1.61$ for LIF.
For ATA-T, the corresponding values are $0.73/1.46$
and $0.95/2.44$. Across all six configurations, $R^2$
is approximately 0.98 or higher.

Figure~\ref{fig:bram_size_evaluation} compares three
(PRE, CURR) pairs for SYNAPTIC ATA-T. At (256,128),
MAE/RMSE are approximately $0.73/1.46$, with
$R^2 \approx 0.99$. At (32,32), MAE/RMSE are approximately
$1.2/1.3$, with $R^2 \approx 0.88$. Thus, the smaller
configuration has a higher MAE but a slightly lower RMSE.
The largest errors occur at (32,128), where MAE/RMSE
reach approximately $4.9/7.0$ and $R^2$ falls to about
0.39, the lowest of the three cases. Since PRE remains
at 32 between (32,32) and (32,128), this increase in
error comes from raising CURR from 32 to 128.

MAE and RMSE are reported in 36K-equivalent blocks.
All numerical values are approximate readings from the plots.

\par
\end{adjustwidth}

\begin{figure}[!ht]
  \centering
  \begin{subfigure}{\columnwidth}
    \centering
    \includegraphics[width=\linewidth]
      {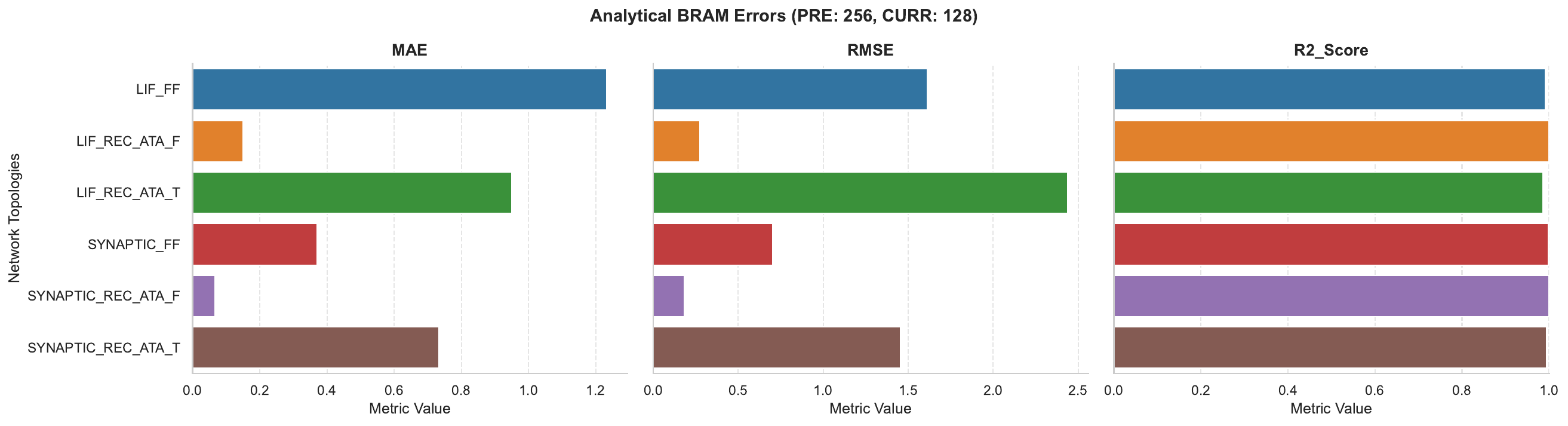}
    \caption{Neuron--topology comparison at PRE = 256
    and CURR = 128.}
    \label{fig:bram_topology_evaluation}
  \end{subfigure}

  \medskip

  \begin{subfigure}{\columnwidth}
    \centering
    \includegraphics[width=\linewidth]
      {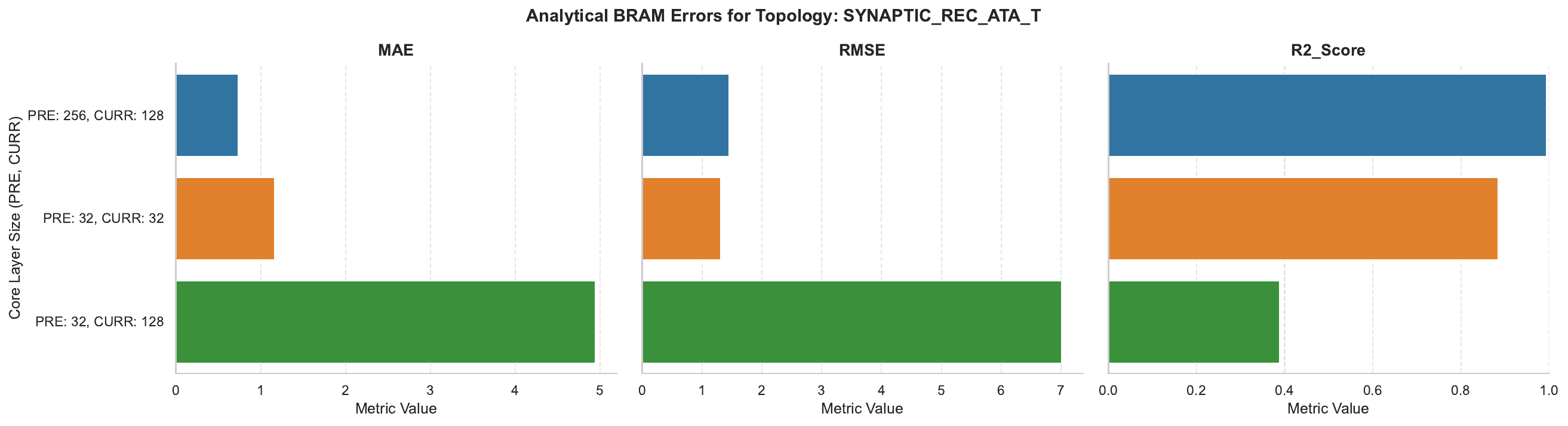}
    \caption{Comparison across layer dimensions for
    SYNAPTIC ATA-T.}
    \label{fig:bram_size_evaluation}
  \end{subfigure}

  \caption{Analytical BRAM estimation accuracy for
  (a) six neuron--topology configurations with fixed
  layer dimensions and (b) three layer-dimension pairs
  for SYNAPTIC ATA-T. Columns report MAE, RMSE, and
  $R^2$, from left to right. MAE and RMSE are expressed
  in 36K-equivalent blocks. PRE and CURR denote the
  previous- and current-layer neuron counts, respectively.}
  \label{fig:bram_evaluation}
\end{figure}

\subsection{Optimization Algorithm and Search Procedure}
\label{sec:optimization_search}

\paragraph{Design Space and Configuration Evaluation}
The search covers the feedforward and recurrent weight
bit-widths, along with the neuron parameters $\beta$ and
$\alpha$. Only parameters relevant to the selected neuron
model are explored; for example, $\alpha$ is excluded
for LIF neurons.

An \emph{evaluated configuration} is one for which hardware
resource usage has been estimated, quantized-inference accuracy
has been measured, and the total cost has been computed from
these two terms. Configurations proposed for evaluation but
not yet assessed are called \emph{candidate configurations}.

Given the size of the design space, an optimization algorithm
chooses which candidate to evaluate next. This choice determines
which configurations are evaluated within the available
evaluation budget.

\paragraph{PSO Search Strategy}
We use particle swarm optimization (PSO) to search for a
low-cost precision configuration that balances classification
accuracy and hardware utilization. Each particle represents
a candidate configuration. Its movement combines its previous
velocity with attraction toward its own best-known position
and the best position found by the swarm. Continuous updates
are mapped to a discrete configuration grid, and random
unseen configurations are introduced to avoid repeated
evaluations.

\paragraph{PSO Search Procedure}
Algorithm~\ref{alg:pso} searches over $(fw,rw,\beta,\alpha)$
with a unique-evaluation budget $M$ determined by
$\rho_{\text{budget}}$ (lines~2--3). Accuracy loss and hardware
cost receive equal weight. Within the hardware term, LUTs,
flip-flops, and BRAMs receive weights of $0.2$, $0.3$, and
$0.5$, respectively (lines~4--5). The swarm has at most
20 particles, further limited by the evaluation budget and
search-space size. Lines~6--11 set the remaining search
controls: the iteration cap, stall patience, inertia range,
attraction coefficients, and velocity limit.

Separate caches avoid repeating accuracy simulations and
hardware estimates (line~12). Hardware estimates are summed
across layers and cached by the bit-width pair $(fw,rw)$
(lines~15--18). Configurations with the same weight precisions
therefore reuse these totals even when $\beta$ or $\alpha$
differs. The totals are normalized by the resource-specific
maxima $\mathbf{h}_{\text{max}}$, with $\epsilon$ preventing
division by zero, and combined into the weighted hardware
penalty (lines~19--20). SNN inference is run only when the
complete configuration has no cached accuracy cost. Its
percentage accuracy is converted to a weighted error cost
and added to the hardware penalty (lines~22--27).

The swarm starts from distinct, evaluated configurations,
each counted toward $M$ (lines~30--35). Particles are
represented by grid-index positions and assigned random
initial velocities. Each starting position becomes the
particle's personal best, and the lowest-cost starting
configuration becomes the global best (lines~36--39).

Search continues within both the unique-evaluation budget
and the iteration limit (lines~42--44). Each iteration
resets the new-evaluation counter and linearly decreases
inertia from $0.90$ toward $0.40$, reducing the influence
of previous motion (lines~45--46). Particles are updated
sequentially, with the remaining budget checked before
each update (lines~47--48). Independent random factors
scale attraction toward the personal and global bests,
and each velocity component is clipped to $[-2.5,2.5]$
(lines~49--51). The resulting position is bounded by
$\mathbf{D}_{\max}$, rounded to the nearest grid point,
and mapped back to a configuration (lines~52--53).

If that configuration has already been evaluated, the
algorithm attempts to substitute a random unseen one.
A successful substitution also resets the particle's
velocity (lines~54--56). Only previously unseen
configurations count toward the evaluation budget;
cached revisits do not (lines~57--59). A particle moves
to the selected configuration even if its cost is higher,
but personal and global bests change only when the cost
strictly decreases (lines~60--62). Sequential updates
allow later particles to use a global best found earlier
in the same iteration.

Stalling is defined by the absence of new evaluations,
not by a lack of cost improvement. Eight consecutive
iterations without a new evaluation terminate the search
early (lines~63--64). The lowest-cost configuration,
its cost, the number of unique evaluations, and the
elapsed time are then recorded in the PSO results sheet
(line~66).

\par\vspace{\baselineskip}

\noindent
\begin{tcolorbox}[
    breakable,
    colback=gray!4,
    colframe=black!85,
    boxrule=0.7pt,
    arc=0pt,
    left=5pt, right=5pt, top=6pt, bottom=6pt
]
\begin{lstlisting}[
    mathescape=true,
    numbers=left,
    numberstyle=\tiny\color{black!60},
    numbersep=5pt,
    basicstyle=\ttfamily\fontsize{8.5pt}{10.4pt}\selectfont,
    lineskip=0.6pt,
    keywordstyle=\bfseries\color{black},
    commentstyle=\itshape\color{black!65},
    escapeinside={(*@}{@*)},
    xleftmargin=14pt
]
# --- Search Space & Swarm Hyperparameters ---
$\mathcal{S} \leftarrow \mathcal{W}_{\text{fw}} \times \mathcal{W}_{\text{rw}} \times \mathcal{B}_{\beta} \times \mathcal{B}_{\alpha}$ (*@\hfill@*) # Discrete search space
$M \leftarrow \min(|\mathcal{S}|, \lceil \rho_{\text{budget}} |\mathcal{S}| \rceil)$ (*@\hfill@*) # Unique-evaluation budget
$w_{\text{acc}}, w_{\text{hw}} \leftarrow 0.5, 0.5$ (*@\hfill@*) # Accuracy vs. hardware weight
$w_{\text{lut}}, w_{\text{ff}}, w_{\text{bram}} \leftarrow 0.2, 0.3, 0.5$ (*@\hfill@*) # LUT, FF, BRAM weights
$N_p \leftarrow \min(20, M, |\mathcal{S}|)$ (*@\hfill@*) # Swarm size
$K_{\max} \leftarrow \max(10, 5 \lceil M / N_p \rceil)$ (*@\hfill@*) # Iteration cap
$\tau_{\text{stall}} \leftarrow 8$ (*@\hfill@*) # Stall patience (iterations)
$\omega_{\max}, \omega_{\min} \leftarrow 0.90, 0.40$ (*@\hfill@*) # Inertia weight range
$c_1, c_2 \leftarrow 1.70, 1.70$ (*@\hfill@*) # Cognitive and social factors
$v_{\max} \leftarrow 2.5$ (*@\hfill@*) # Velocity clamp
$\text{acc\_cache}, \text{hw\_cache} \leftarrow \{\}, \{\}$ (*@\hfill@*) # Memoization caches

# --- Layer-wise Hardware Estimation & Configuration Evaluation ---
$\text{estimate\_hw}(fw, rw)$:
  $k_{\text{hw}} \leftarrow (fw, rw)$ (*@\hfill@*) # Cache key from bit-widths
  if $k_{\text{hw}} \notin \text{hw\_cache}$: (*@\hfill@*) # Cache miss
    $\text{hw\_cache}[k_{\text{hw}}] \leftarrow \sum_{\ell} \text{predict\_layer}(\ell, fw, rw)$ (*@\hfill@*) # Sum resources over layers
  $\bar{h}_{\text{lut}}, \bar{h}_{\text{ff}}, \bar{h}_{\text{bram}} \leftarrow \text{hw\_cache}[k_{\text{hw}}]_{\text{raw}} / (\mathbf{h}_{\text{max}} + \epsilon)$ (*@\hfill@*) # Normalize by maxima
  return $w_{\text{hw}} (w_{\text{lut}}\bar{h}_{\text{lut}} + w_{\text{ff}}\bar{h}_{\text{ff}} + w_{\text{bram}}\bar{h}_{\text{bram}})$ (*@\hfill@*) # Weighted hardware penalty

$\text{evaluate\_config}(s)$:
  $fw, rw, \beta, \alpha \leftarrow s$ (*@\hfill@*) # Unpack configuration
  if $s \notin \text{acc\_cache}$: (*@\hfill@*) # Cache miss only
    $\text{acc} \leftarrow \text{run\_snn\_exact\_inference}(s)$ (*@\hfill@*) # Exact SNN inference
    $\text{acc\_cache}[s] \leftarrow (1.0 - \text{acc}/100.0) \times w_{\text{acc}}$ (*@\hfill@*) # Weighted accuracy cost
  return $\text{acc\_cache}[s] + \text{estimate\_hw}(fw, rw)$ (*@\hfill@*) # Total cost

# --- Swarm Initialization ---
$\{s_i\}_{i=1}^{N_p} \leftarrow \text{sample\_unique}(\mathcal{S}, N_p)$ (*@\hfill@*) # Random distinct initial states
$\text{eval\_count} \leftarrow 0$;  $C_{\text{gbest}} \leftarrow \infty$ (*@\hfill@*) # Counter and global best cost
for $i = 1$ to $N_p$: (*@\hfill@*) # Initialize each particle
  if $s_i \notin \text{acc\_cache}$: (*@\hfill@*) # Unseen state?
    $\text{eval\_count} \leftarrow \text{eval\_count} + 1$ (*@\hfill@*) # Count new evaluation
  $C_i \leftarrow \text{evaluate\_config}(s_i)$ (*@\hfill@*) # Initial cost
  $\mathbf{x}_i \leftarrow \text{to\_pos}(s_i)$ (*@\hfill@*) # Grid position
  $\mathbf{v}_i \sim \mathcal{U}(-v_{\max}, v_{\max})^4$ (*@\hfill@*) # Random initial velocity
  $s_{\text{pbest},i}, \mathbf{x}_{\text{pbest},i}, C_{\text{pbest},i} \leftarrow s_i, \mathbf{x}_i, C_i$ (*@\hfill@*) # Personal best = start
  if $C_i < C_{\text{gbest}}$ then $s^*, \mathbf{x}_{\text{gbest}}, C_{\text{gbest}} \leftarrow s_i, \mathbf{x}_i, C_i$ (*@\hfill@*) # Update global best

# --- Main Discrete PSO Optimization Loop ---
$k \leftarrow 0$;  $n_{\text{stall}} \leftarrow 0$ (*@\hfill@*) # Iteration and stall counters
while ($\text{eval\_count} < M$) and ($k < K_{\max}$): (*@\hfill@*) # Budget or iteration cap
  $k \leftarrow k + 1$ (*@\hfill@*) # Next iteration
  $n_{\text{new}} \leftarrow 0$ (*@\hfill@*) # New evaluations this iteration
  $\omega \leftarrow \omega_{\max} - (\omega_{\max} - \omega_{\min}) \frac{k - 1}{\max(1, K_{\max} - 1)}$ (*@\hfill@*) # Linear inertia decay
  for $i = 1$ to $N_p$: (*@\hfill@*) # Sequential particle sweep
    if $\text{eval\_count} \geq M$ then break (*@\hfill@*) # Stop at budget
    $\mathbf{r}_1, \mathbf{r}_2 \sim \mathcal{U}(0, 1)^4$ (*@\hfill@*) # Random factors per dimension
    $\Delta \mathbf{v}_i \leftarrow c_1 \mathbf{r}_1 \odot (\mathbf{x}_{\text{pbest},i} - \mathbf{x}_i) + c_2 \mathbf{r}_2 \odot (\mathbf{x}_{\text{gbest}} - \mathbf{x}_i)$ (*@\hfill@*) # Pull toward global best
    $\mathbf{v}_i \leftarrow \text{clip}(\omega \mathbf{v}_i + \Delta \mathbf{v}_i, -v_{\max}, v_{\max})$ (*@\hfill@*) # Clamped velocity update
    $\hat{\mathbf{x}}_i \leftarrow \text{clip}(\mathbf{x}_i + \mathbf{v}_i, \mathbf{0}, \mathbf{D}_{\max})$ (*@\hfill@*) # Continuous position update
    $s' \leftarrow \text{to\_state}(\lfloor \hat{\mathbf{x}}_i + 0.5 \rfloor)$ (*@\hfill@*) # Snap to nearest grid state
    if $s' \in \text{acc\_cache}$: (*@\hfill@*) # Duplicate proposal
      $s_{\text{imm}} \leftarrow \text{random\_unseen}(\mathcal{S}, \text{acc\_cache})$ (*@\hfill@*) # Draw unseen state
      if $s_{\text{imm}} \neq \text{None}$ then $s' \leftarrow s_{\text{imm}}$;  $\mathbf{v}_i \leftarrow \mathbf{0}$ (*@\hfill@*) # Immigrant; reset velocity
    if $s' \notin \text{acc\_cache}$: (*@\hfill@*) # Unseen state
      $\text{eval\_count} \leftarrow \text{eval\_count} + 1$;  $n_{\text{new}} \leftarrow n_{\text{new}} + 1$ (*@\hfill@*) # Count new evaluation
    $C_i \leftarrow \text{evaluate\_config}(s')$ (*@\hfill@*) # Cost of new state
    $s_i, \mathbf{x}_i \leftarrow s', \text{to\_pos}(s')$ (*@\hfill@*) # Move particle
    if $C_i < C_{\text{pbest},i}$ then $s_{\text{pbest},i}, \mathbf{x}_{\text{pbest},i}, C_{\text{pbest},i} \leftarrow s_i, \mathbf{x}_i, C_i$ (*@\hfill@*) # Update personal best
    if $C_i < C_{\text{gbest}}$ then $s^*, \mathbf{x}_{\text{gbest}}, C_{\text{gbest}} \leftarrow s_i, \mathbf{x}_i, C_i$ (*@\hfill@*) # Update global best
  if $n_{\text{new}} = 0$ then $n_{\text{stall}} \leftarrow n_{\text{stall}} + 1$ else $n_{\text{stall}} \leftarrow 0$ (*@\hfill@*) # Count stalled iterations
  if $n_{\text{stall}} \geq \tau_{\text{stall}}$ then break (*@\hfill@*) # Early stop

$\text{append\_best\_to\_excel\_pso}(s^*, C_{\text{gbest}}, \text{eval\_count}, t_{\text{elapsed}})$ (*@\hfill@*) # Write best row to PSO sheet
\end{lstlisting}
\end{tcolorbox}\par
\vspace{4pt}

\refstepcounter{lstlisting}\label{alg:pso}
\noindent
\begin{minipage}{\linewidth}
    \fontsize{8.8pt}{11pt}\selectfont
    \rmfamily\textbf{Algorithm~\thelstlisting:} Precision-aware SNN discrete particle swarm optimization (PSO) DSE with cache-assisted random immigrant injection and stall-aware budget control. Continuous index-space trajectories are quantized onto the 4D discrete grid $(fw, rw, \beta, \alpha)$, while memoization caches prevent redundant inference and hardware evaluations.
\end{minipage}
\vspace{8pt}

\subsection{RTL Configurator}
\label{subsec:rtl_configurator}

After DSE, the RTL configurator combines the lowest-cost precision
configuration with the network structure and other parameters
specified through the GUI. It sets the Flexi-NeurA compile-time
parameters for the selected topology, layer sizes, numerical
precisions, and neuron parameters, without requiring manual
changes to the RTL.

The resulting package contains the configured HDL design,
hardware drivers, execution and validation software, quantized
weights and neuron parameters, and the encoded input and
parameter data needed to run and evaluate the accelerator.

\section{Results and Evaluation}\label{sec:results}

\subsection{Case Studies: Auditory, Gesture, and Vision Processing}\label{sec:case_study}

To demonstrate the practical relevance of Flexi-NeurA for diverse edge-AI applications, we present comprehensive, end-to-end evaluations across three distinct domains: auditory processing using the Spiking Heidelberg Digits (SHD) dataset~\cite{ref30}, dynamic vision sensor (DVS) gesture recognition~\cite{ref28}, and standard vision classification using the MNIST benchmark~\cite{ref29}.

We evaluate Flexi-NeurA end-to-end on standard neuromorphic benchmarks using the Flex-Plorer framework and synthesized FPGA implementations. As mentioned in Section~\ref{sec:Flexi-NeurA}, each network layer is mapped to a dedicated processing core, and inter-layer communication is realized exclusively through address-event packets. Inter-core event transfer is handled through a handshake-based protocol that provides backpressure and buffering, guaranteeing lossless communication.

\begin{figure*}[htbp]
  \centering
  \includegraphics[width=\linewidth]{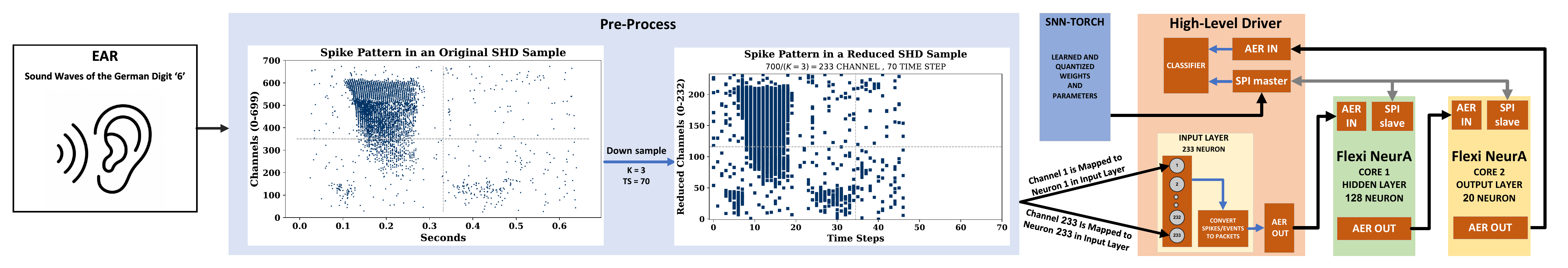}
  \caption{End-to-end data flow and hardware mapping for the SHD auditory processing case study.}
  \label{fig:flow_shd}
\end{figure*}

\begin{figure*}[htbp]
  \centering
  \includegraphics[width=\linewidth]{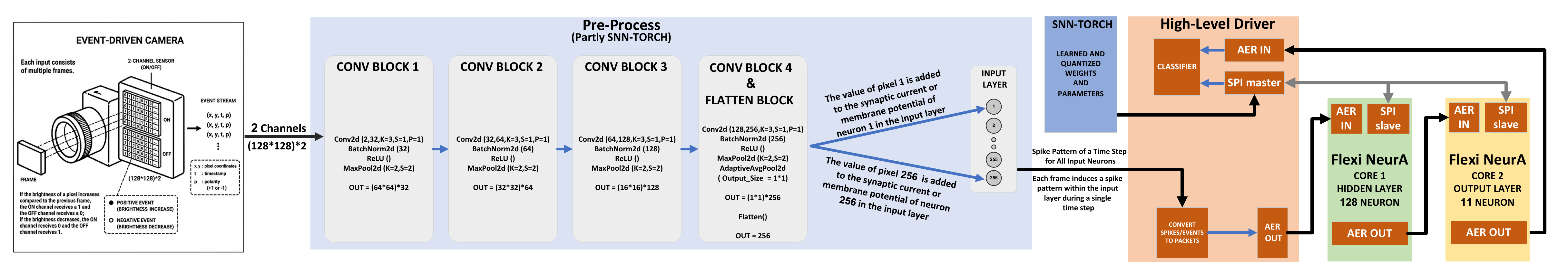}
  \caption{End-to-end data flow and hardware mapping for DVS-based gesture processing.}
  \label{fig:flow_dvs}
\end{figure*}

For SHD auditory processing, as shown in
Fig.~\ref{fig:flow_shd}, the dataset provides spike streams
distributed across $700$ frequency channels with continuous-valued
timestamps. To accommodate the hardware constraints, preprocessing
reduces the input dimensionality to $233$ channels and discretizes
each sample into $70$ time steps. A software driver maps each
input channel to a corresponding input neuron. At each time step,
the driver scans the $233$ input channels, encodes the input spikes
as AER packets, and transmits them to the hardware. Flexi-NeurA is
configured via SPI using quantized weights extracted from the
network trained in snnTorch~\cite{ref14}. Core~1 implements a
$128$-neuron hidden layer and forwards its output spikes to
Core~2, which implements a $20$-neuron output layer for
spoken-digit classification.

For DVS gesture recognition (Fig.~\ref{fig:flow_dvs}), the event
camera records pixel-level brightness changes as asynchronous
events, each specifying a pixel location, timestamp, and polarity.
Preprocessing converts these events into a sequence of
$128 \times 128$ tensors with two channels representing ON and
OFF polarities. Each tensor passes through a convolutional
preprocessing pipeline comprising four spatial-reduction blocks
and a flattening operation, producing a $256$-element feature
vector. Each feature is mapped to a corresponding input neuron
and contributes to either its synaptic current or membrane
potential, depending on the neuron model. At each discrete time
step, the input neurons generate spikes according to their
states and update rules. These spikes are encoded as AER packets
and transmitted to Core~1, which implements a $128$-neuron hidden
layer. Its output spikes are then forwarded to Core~2, which
implements an $11$-neuron output layer for gesture classification.

For MNIST digit classification, as illustrated in
Fig.~\ref{fig:multicore}, grayscale images of $28 \times 28$
pixels are downsampled to $16 \times 16$ pixels and normalized.
The pixel values are converted into spike trains using rate
encoding over a fixed inference window, yielding $256$ input
channels. The resulting spikes are encoded as AER packets and
transmitted to the processing cores. The SNNs are trained offline
using backpropagation through time (BPTT) with surrogate gradients
in snnTorch~\cite{ref14}. Following training, the synaptic weights
and the decay coefficients applicable to the selected neuron
model are quantized and exported for hardware deployment.

\subsection{Flexi-NeurA Evaluation and Results}

Flexi-NeurA supports three topologies combined with three neuron models, yielding the design space reported in Table~\ref{tab:combined_experiments} from hardware profiling on SHD, DVS, and MNIST. No single neuron model or topology is inherently superior to another; rather, given the diverse and heterogeneous nature of input signal spaces, the choice is application-dependent, determined by which neuron--topology combination best matches the target input domain and yields the highest performance for it.

Readout mechanism selection is causally tied to task complexity and network depth. On SHD---a temporally rich, biologically realistic dataset---potential-based readout consistently outperforms spike-rate readout across all neuron models (IF, LIF, Synaptic) and topologies, in both the training and hardware-aware inference phases, owing to its preservation of fine-grained temporal dynamics. On DVS and MNIST, this advantage is confined to the ATA-T topology, where deep recurrent feedback amplifies the benefit of potential-based readout; under FF and ATA-F configurations, both readout types yield comparable accuracy. These results establish that potential-based readout is essential when dataset biological realism is high or recurrent temporal depth is significant, while spike-rate readout offers a competitive, lower-cost alternative in simpler configurations.

\begin{table}[htbp]
  \centering
 \caption{Hardware/software accuracy (HW/SW, \%) and energy
(En., mJ) across configurations. Accuracy values are rounded
to the nearest integer. Precision and time steps ($T$) are
fixed per dataset.}

  \label{tab:combined_experiments}
  \resizebox{\linewidth}{!}{%
    \begin{tabular}{@{}llcccccc@{}}
      \toprule
      \multirow{2}{*}{\textbf{Dataset (Prec., $T$)}}
      & \multirow{2}{*}{\textbf{Neuron}}
      & \multicolumn{2}{c}{\textbf{FF}}
      & \multicolumn{2}{c}{\textbf{ATA-F}}
      & \multicolumn{2}{c}{\textbf{ATA-T}} \\
      \cmidrule(lr){3-4}
      \cmidrule(lr){5-6}
      \cmidrule(l){7-8}
      & & \textbf{HW/SW} & \textbf{En.}
        & \textbf{HW/SW} & \textbf{En.}
        & \textbf{HW/SW} & \textbf{En.} \\
      \midrule

      \multirow{4}{*}{%
        \begin{tabular}[c]{@{}l@{}}
          \textbf{SHD} \\
          (10-bit, $T=70$)
        \end{tabular}}
      & \textit{Readout:}
        & \multicolumn{2}{c}{\textit{Potential}}
        & \multicolumn{2}{c}{\textit{Potential}}
        & \multicolumn{2}{c}{\textit{Potential}} \\
      & IF       & 73/74 & 1.48 & 66/69 & 1.85 & 61/70 & 2.77 \\
      & LIF      & 75/76 & 1.48 & 70/72 & 1.85 & 65/75 & 2.77 \\
      & SYNAPTIC & 71/72 & 1.66 & 66/76 & 2.03 & 66/74 & 3.33 \\
      \midrule

      \multirow{4}{*}{%
        \begin{tabular}[c]{@{}l@{}}
          \textbf{DVS} \\
          (8-bit, $T=80$)
        \end{tabular}}
      & \textit{Readout:}
        & \multicolumn{2}{c}{\textit{Spike Rate}}
        & \multicolumn{2}{c}{\textit{Spike Rate}}
        & \multicolumn{2}{c}{\textit{Potential}} \\
      & IF       & 93/93 & 0.55 & 89/91 & 0.66 & 86/92 & 1.29 \\
      & LIF      & 93/91 & 0.64 & 93/93 & 1.11 & 88/93 & 1.85 \\
      & SYNAPTIC & 87/89 & 2.59 & 86/93 & 3.14 & 84/92 & 5.18 \\
      \midrule

      \multirow{4}{*}{%
        \begin{tabular}[c]{@{}l@{}}
          \textbf{MNIST} \\
          (6-bit, $T=100$)
        \end{tabular}}
      & \textit{Readout:}
        & \multicolumn{2}{c}{\textit{Spike Rate}}
        & \multicolumn{2}{c}{\textit{Spike Rate}}
        & \multicolumn{2}{c}{\textit{Potential}} \\
      & IF       & 97/97 & 0.11 & 95/95 & 0.12 & 96/97 & 0.25 \\
      & LIF      & 97/97 & 0.12 & 96/97 & 0.14 & 94/97 & 0.37 \\
      & SYNAPTIC & 97/97 & 0.14 & 97/97 & 0.24 & 95/98 & 0.37 \\
      \bottomrule
    \end{tabular}%
  }
\end{table}

As shown in Table~\ref{tab:combined_experiments}, the evaluations across SHD, DVS, and MNIST expose a consistent trade-off between recurrence and efficiency. On SHD, LIF-FF is the most accurate configuration at $75\%$ for $1.48$~mJ, while its ATA-T counterpart loses $10$~percentage points ($65\%$) and consumes $87\%$ more energy ($2.77$~mJ); the SYNAPTIC neuron marks the highest cost on this dataset with $3.33$~mJ in ATA-T at $66\%$ accuracy. On DVS, IF-FF and LIF-FF both reach $93\%$ at only $0.55$~mJ and $0.64$~mJ, whereas their ATA-T variants drop to $86\%$ and $88\%$ at $1.29$~mJ and $1.85$~mJ, respectively; SYNAPTIC is the least efficient model overall, requiring $2.59$~mJ in FF and $5.18$~mJ in ATA-T for $87\%$ and $84\%$ accuracy. On MNIST, every configuration stays within $94\%$--$97\%$ accuracy at a minimal budget of $0.11$--$0.37$~mJ, so the relative penalty of recurrence becomes most visible on this dataset: energy grows by only $9\%$ for IF-ATA-F ($0.11\to0.12$~mJ) but by up to $208\%$ for LIF-ATA-T ($0.12\to0.37$~mJ), while the largest MNIST accuracy loss is $3$~percentage points ($97\%\to94\%$ for LIF-ATA-T). Across all datasets, adding the ATA-F and ATA-T topologies increases energy consumption without exception, although the magnitude of the overhead is dataset- and neuron-dependent: relative to the matching FF baseline it ranges from $+22\%$ to $+101\%$ on SHD, from $+20\%$ to $+189\%$ on DVS, and from $+9\%$ to $+208\%$ on MNIST. Accuracy, in turn, never improves: it either stays unchanged (e.g., LIF on DVS and SYNAPTIC on MNIST, both at $93\%$ and $97\%$ in FF and ATA-F) or degrades by up to $12$~percentage points (IF on SHD, $73\%\to61\%$).

\paragraph{Comparison with State-of-the-Art}\label{sec:baseline_results}

Table~\ref{tab:comparison} compares Flexi-NeurA with recent FPGA-based accelerators on the MNIST benchmark. SNN accelerators prioritize hardware efficiency. Within this class, Flexi-NeurA achieves the second-highest accuracy, reaching $96.68\%$ on MNIST with a marginal difference from the state-of-the-art.

\begin{table*}[t]
  \caption{Comparison of Flexi-NeurA with recent FPGA-based accelerators on the MNIST benchmark.}
  \label{tab:comparison}
  \centering
  \renewcommand{\arraystretch}{1.2}
  \setlength{\tabcolsep}{3.5pt}
  \resizebox{\textwidth}{!}{%
    \begin{tabular}{@{}llccccccccccccccc@{}}
      \toprule
      \multirow{2}{*}{\begin{tabular}{@{}l@{}}Design [Ref] \\ (Year)\end{tabular}} &
      \multirow{2}{*}{FPGA} &
      \multirow{2}{*}{\begin{tabular}{@{}c@{}}Model \\ (Update)\end{tabular}} &
      \multirow{2}{*}{\begin{tabular}{@{}c@{}}W/N \\ (bits)\end{tabular}} &
      \multirow{2}{*}{\begin{tabular}{@{}c@{}}$f_{\mathrm{clk}}$ \\ (MHz)\end{tabular}} &
      \multicolumn{3}{c}{Hardware (Used/Avail.)} &
      \multicolumn{2}{c}{Topology} &
      \multirow{2}{*}{$N_{\mathrm{ts}}$} &
      \multicolumn{4}{c}{Performance} &
      \multirow{2}{*}{\begin{tabular}{@{}c@{}}Acc. \\ (\%)\end{tabular}} \\
      \cmidrule(lr){6-8} \cmidrule(lr){9-10} \cmidrule(lr){12-15}
      & & & & & BRAM & DSP & Logic Cell & Architecture & \#Syn & & \begin{tabular}{@{}c@{}}$T_{\mathrm{lat}}$ \\ (ms)\end{tabular} & \begin{tabular}{@{}c@{}}Power \\ (W)\end{tabular} & \begin{tabular}{@{}c@{}}$E_{\mathrm{img}}$ \\ (mJ)\end{tabular} & \begin{tabular}{@{}c@{}}$E_{\mathrm{syn}}$ \\ (nJ)\end{tabular} & \\
      \midrule
      Han et al.~\cite{ref21} (2020) & XC7Z045 & LIF (Evt) & 16/16 & 200 & 40.5/545 & 0/900 & 12.7k/655.8k & \begin{tabular}{@{}c@{}}784-1024-\\1024-10\end{tabular} & 1.86M & N/R & 6.21 & 0.477 & 2.96 & 1.59 & 97.06 \\
      Gupta et al.~\cite{ref24} (2020) & XC6VLX240T & LIF (Evt) & 24/24 & 100 & 162/416 & 64/768 & 79.5k/452.2k & 784-16 & 12.5k & N/R & 0.50 & N/R & N/R & N/R & N/R \\
      Li et al.~\cite{ref22} (2021) & XC7VX485 & LIF (Hyb) & 16/16 & 100 & N/R/2060 & N/R/2800 & N/R/485.8k & \begin{tabular}{@{}c@{}}784-200-\\100-10\end{tabular} & 178k & N/R & 3.15 & 1.6 & 5.04 & 28 & 92.93 \\
      Garpegna et al.~\cite{ref25} (2022) & XC7Z020 & LIF (Clk) & 16/16 & 100 & 45/140 & 0/200 & 56.0k/159.6k & 784-400 & 314k & 3500 & 0.22 & 59.09 & 13 & 41 & 73.96 \\
      SPIKER+~\cite{ref8} (2024) & XC7Z020 & LIF (Clk) & 4/6 & 100 & 18/140 & 0/220 & 7.6k/159.6k & 784-128-10 & 102k & 100 & 0.78 & 0.18 & 0.14 & 1.37 & 93.85 \\
      \midrule
      \textbf{Flexi-NeurA (ours)} & \textbf{XC7Z020} & \textbf{LIF (Evt)} & \textbf{6/8} & \textbf{60} & \textbf{7/140} & \textbf{0/220} & \textbf{1.6k/159.6k} & \textbf{256-128-10} & \textbf{34.0k} & \textbf{100} & \textbf{1.10} & \textbf{0.111} & \textbf{0.12} & \textbf{3.5} & \textbf{96.68} \\
      \bottomrule
    \end{tabular}%
  }
  \\[1ex]
  \raggedright
  \footnotesize{N/R: not reported \quad $N_{\mathrm{ts}}$: number of time steps \quad W/N: weight/neuron bit-width \quad (k: $10^3$, M: $10^6$).}
\end{table*}

On a Xilinx Zynq-7000 XC7Z020 FPGA, the selected MNIST configuration implements a three-layer fully connected SNN with LIF neurons ($256$--$128$--$10$) and $6$-bit quantized weights, mapped to two processing cores. Core~1 hosts the hidden layer, while Core~2 implements the output layer. Post-synthesis results obtained using Vivado show a combined footprint of $1{,}623$ logic cells, $934$ LUTs, $689$ flip-flops, and $7$ BRAM blocks, corresponding to approximately $1\%$, $1.8\%$, $0.6\%$, and $5\%$ of device resources, respectively. Total power consumption is measured at $111~\mathrm{mW}$, dominated by static power.

Flexi-NeurA classifies a single MNIST image in $1.1~\mathrm{ms}$. Despite limited parallelism, latency remains competitive due to the event-driven execution model, which eliminates idle computation and unnecessary memory accesses. The $60~\mathrm{MHz}$ clock frequency, the $100$ simulation time steps per image, and the per-neuron sequential updates determine the overall delay.

With a fixed number of cores and fixed arithmetic precision,
increasing layer width primarily increases memory requirements.
Since neurons within each layer share a time-multiplexed datapath,
the growth of datapath logic is limited. However, additional
addressing and control logic may be required as layer width
increases.

\subsection{Flex-Plorer Evaluation and Results}

This section evaluates Flex-Plorer for precision-aware SNN
design-space exploration. We first compare five non-exhaustive
search methods with exhaustive search across nine DVS design
spaces, using the same evaluation-budget rule for all
non-exhaustive methods. We then examine the hardware--accuracy
trade-offs and selected precision settings in separate DVS
and MNIST examples.

\paragraph{Evaluation Setup}

The search-method comparison uses a $256$--$128$--$11$ SNN
trained on the DVS dataset, with Synaptic neurons and ATA-T
connectivity. The nine design spaces contain approximately
$250$, $500$, $1000$, $1500$, $2000$, $2500$, $3000$, $3500$,
and $4000$ configurations. We compare Bayesian Optimization
(BO), Simulated Annealing (SA), Particle Swarm Optimization
(PSO), Random Search (RS), and Genetic Algorithm (GA).

Each non-exhaustive method receives a budget of one tenth
of the configurations in a given space, rounded up to the
nearest integer. This budget includes initialization and
counts only unique evaluations; revisiting a cached
configuration does not consume additional budget. Exhaustive
Search (EXH) evaluates every configuration to establish the
minimum-cost reference. Reported costs and exploration times
are averaged over ten runs with different random seeds.

\begin{figure*}[t]
  \centering
  \begin{subfigure}[b]{0.48\textwidth}
    \centering
    \includegraphics[width=\linewidth]{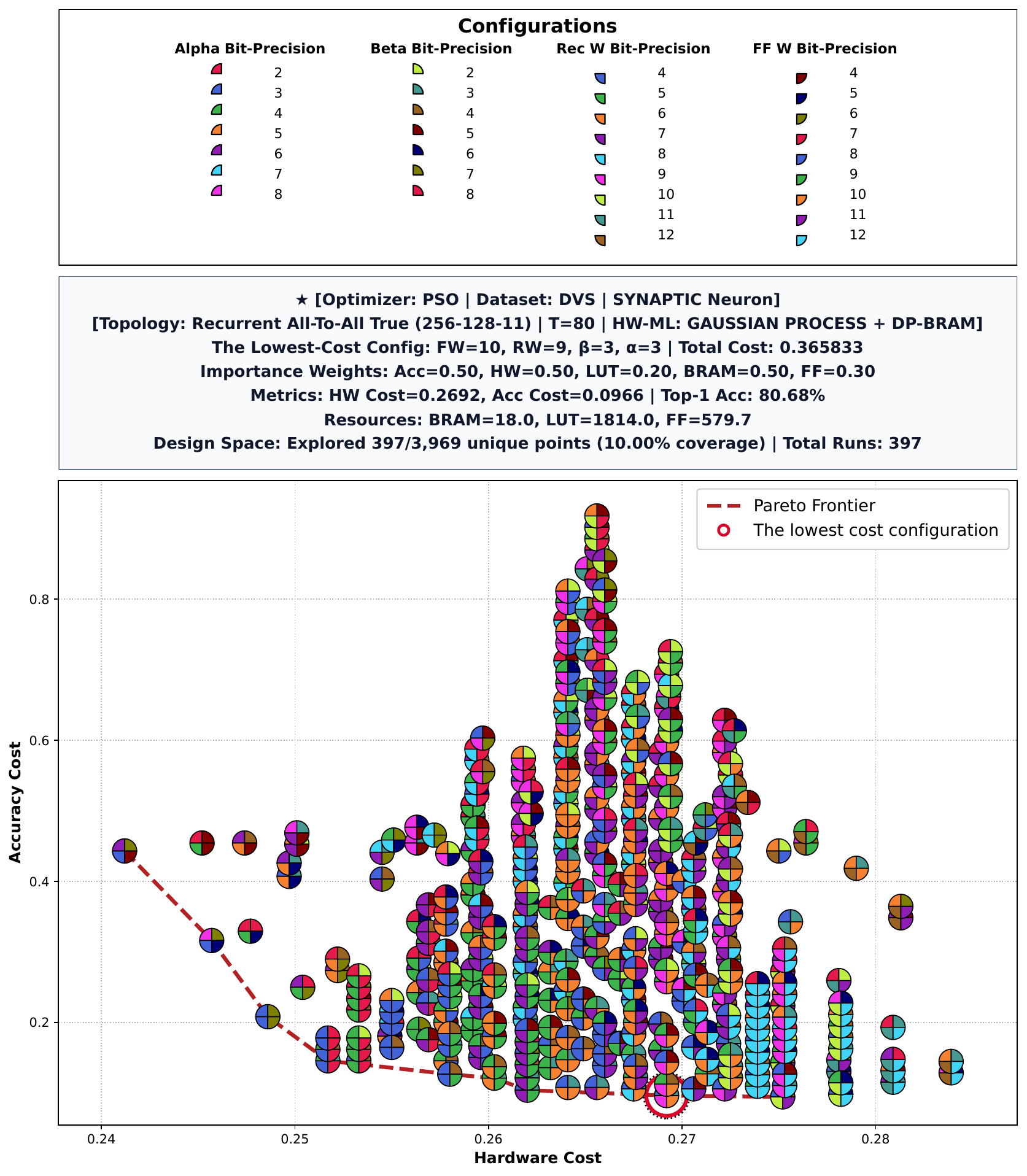}
    \caption{DVS gesture recognition}
    \label{fig:dse_dvs}
  \end{subfigure}
  \hfill
  \begin{subfigure}[b]{0.48\textwidth}
    \centering
    \includegraphics[width=\linewidth]{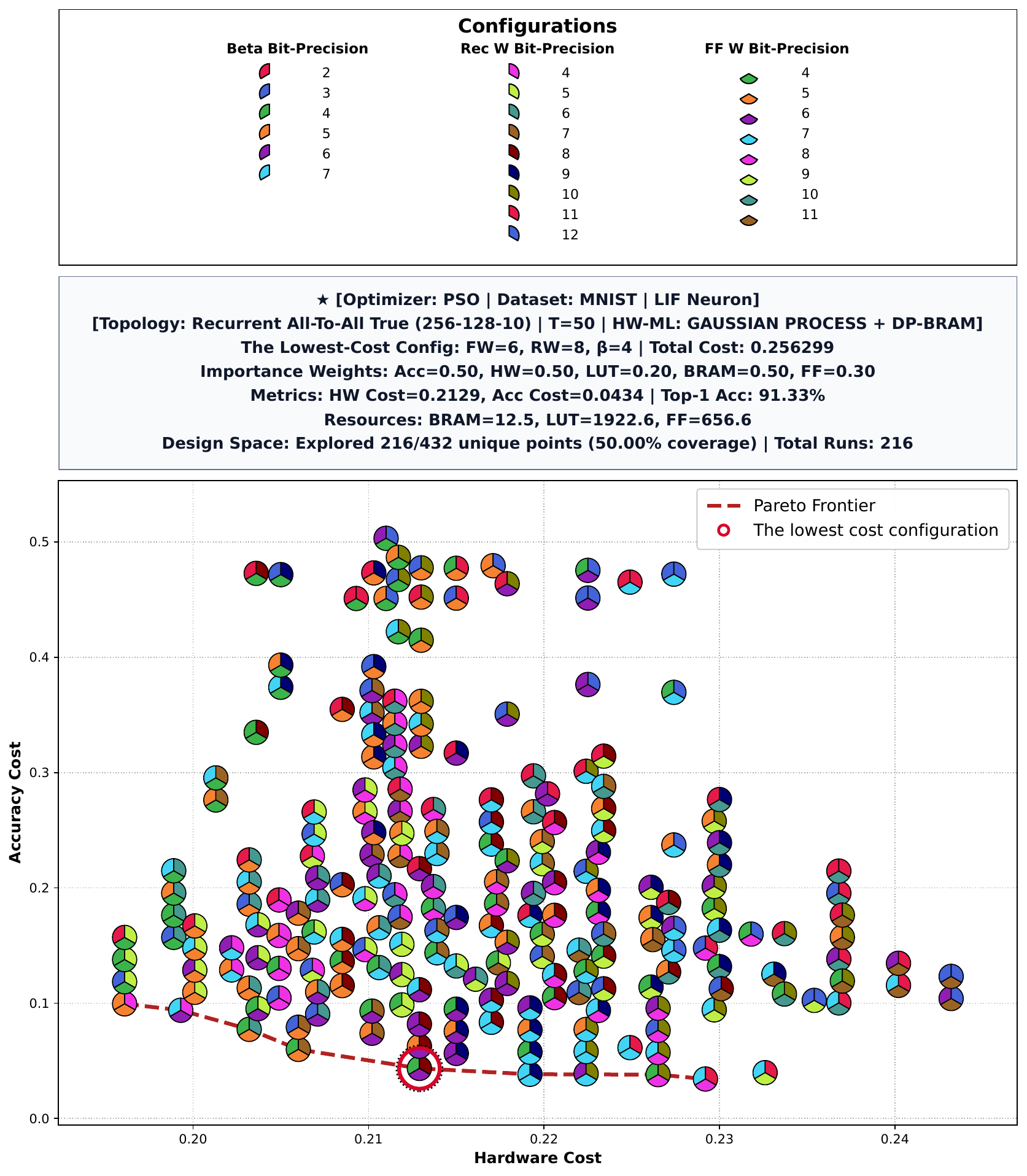}
    \caption{MNIST classification}
    \label{fig:dse_mnist}
  \end{subfigure}
  \caption{Hardware--accuracy trade-offs for evaluated
  configurations of (a) DVS and (b) MNIST. Colored segments
  encode parameter bit-widths, red dashed lines connect
  nondominated evaluated configurations, and red circles
  mark the lowest-total-cost configurations among those
  evaluated. Boxed annotations summarize the search settings
  and results.}
  \label{fig:dse_combined}
\end{figure*}

The separate DVS and MNIST examples illustrate how exploration
balances hardware cost and accuracy when selecting precision
settings. They use the configurations and search coverage
reported in Figure~\ref{fig:dse_combined}, rather than the
common budget rule used in the search-method comparison.

\paragraph{Search Strategies}

All methods use the total cost in \eqref{eq:total_cost},
with accuracy and hardware costs given by \eqref{eq:acc_cost}
and \eqref{eq:hw_cost}, respectively, and weights satisfying
\eqref{eq:constraints}. The methods differ in how they choose
configurations to evaluate. Each selected configuration
undergoes quantized inference and hardware estimation unless
the required results are already cached. The resulting cost
updates the best solution found so far and, for guided methods,
informs subsequent candidate selection. Non-exhaustive searches
stop when they exhaust their budget or meet a method-specific
stopping condition; EXH evaluates the full space.

BO fits a surrogate model to previously evaluated
configurations. Its acquisition function uses predicted
costs and uncertainty to balance the search between promising
and less-explored regions.

SA evaluates a neighbor of the current configuration.
It accepts cost reductions and may also accept cost increases
with a temperature-dependent probability. As the temperature
falls, the search becomes more selective.

PSO represents candidate configurations as particles whose
movement depends on their own best-known solutions and the
swarm's best solution. Updated positions are mapped to valid
discrete configurations before evaluation.

RS samples configurations randomly, without using previous
costs to guide the search. It retains the best configuration
found and serves as a baseline for assessing guided methods.

GA evolves a population using evaluated costs to guide parent
selection. Crossover combines parent configurations, while
mutation introduces new precision combinations.

EXH returns the lowest-cost configuration in the entire
discrete space. It therefore establishes the optimum within
that space, at the expense of more evaluations than the
budget-limited methods.

\paragraph{Search-Method Comparison and Trade-off Analysis}

\begin{figure}[!t]
  \centering
  \includegraphics[width=\columnwidth]{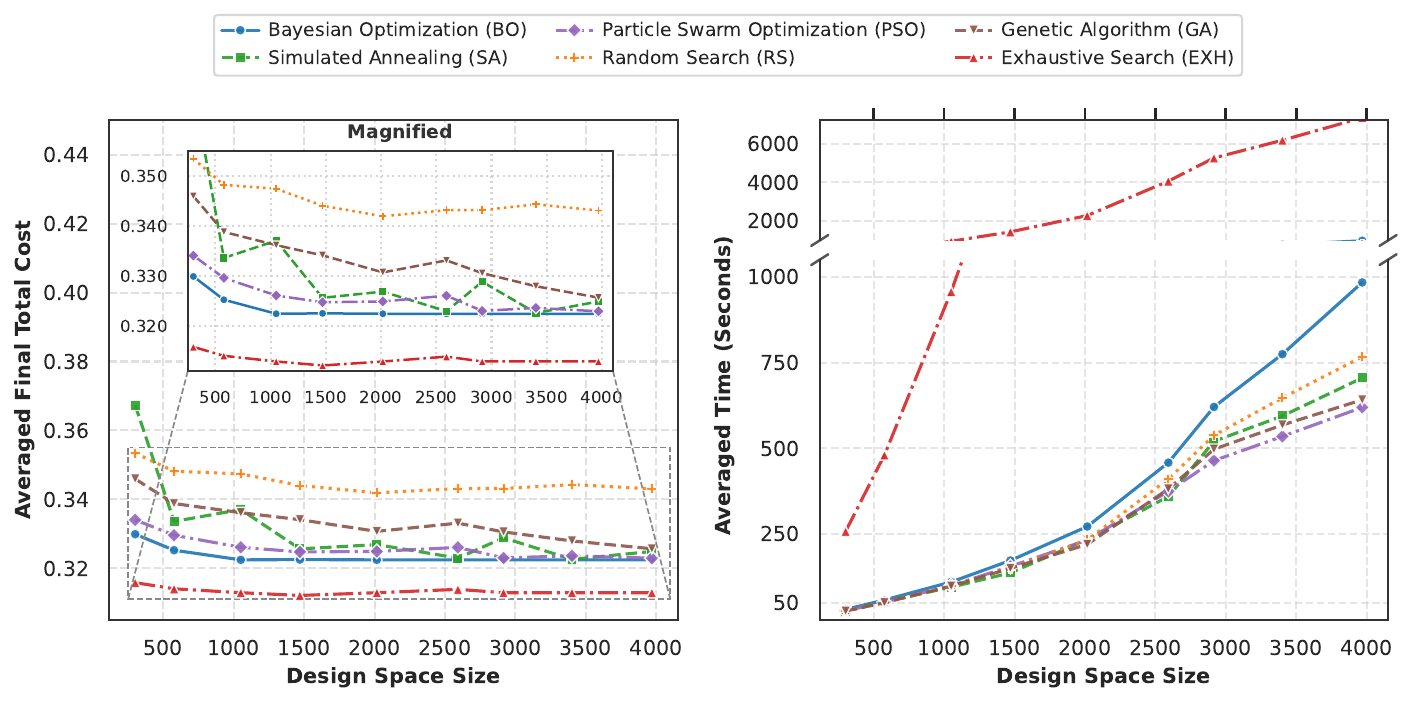}
  \caption{Comparison of search methods on DVS: mean best-found
  total cost (left) and mean exploration time (right) over ten
  runs with different random seeds. Non-exhaustive methods
  receive a budget of ten percent of each design space,
  rounded up, counting only unique evaluations; EXH evaluates
  all configurations. The inset enlarges the low-cost region,
  and the broken time axis accommodates the EXH runtime.}
  \label{fig:exploration_results}
\end{figure}

Figure~\ref{fig:exploration_results} compares the mean
best-found total cost and mean exploration time on DVS.
In the smallest space, SA has a substantially higher mean
cost, approximately $0.439$, than the other non-exhaustive
methods, which achieve $0.320$--$0.324$. The EXH reference
is approximately $0.317$. As the spaces grow, most methods
approach the EXH reference of approximately $0.312$.
GA remains close to this reference from about $500$
configurations onward, while BO and PSO achieve near-reference
costs at several sizes. At the largest size, PSO's cost rises
slightly to approximately $0.315$; SA and RS reach approximately
$0.320$ at some intermediate sizes. These improvements partly
reflect the budget rule, since larger spaces allow more
evaluations under the ten-percent allocation.

Equal evaluation counts do not necessarily lead to equal
runtimes. Exploration time depends on both search overhead
and the configurations evaluated: in the hardware-aware
simulator, larger weight bit-widths take longer to simulate.
Giving accuracy more weight relative to hardware cost may
therefore increase exploration time by favoring higher-precision
configurations, depending on the accuracy gains they provide.

For the largest space, EXH takes approximately $7000$ seconds,
compared with $550$ seconds for SA, $620$ seconds for both GA
and PSO, $670$ seconds for RS, and $980$ seconds for BO.
Although BO achieves some of the lowest costs among the
non-exhaustive methods, it is also the slowest at every
plotted size. Updating its surrogate model and optimizing
the acquisition function add computation to candidate
selection. RS has no such cost-guided selection, but it
may sample high-bit-width configurations that take longer
to simulate. These factors can contribute to the runtime
differences, though the figure does not separate search
overhead from simulation time.

PSO offers a balanced trade-off between solution quality
and runtime, particularly for spaces of approximately
$3000$--$3500$ configurations, where it reaches near-reference
costs in about $370$--$440$ seconds. GA likewise combines
near-reference costs with competitive runtimes. SA is the
fastest method for the largest space, but its costs are
not consistently as low across all sizes. No method is
uniformly superior, and the reported means alone cannot
establish run-to-run reliability or statistical significance.

\paragraph{Hardware--Accuracy Trade-offs}

Figure~\ref{fig:dse_combined} plots the precision configurations
evaluated for DVS and MNIST. Each marker represents one
configuration. Its colored segments indicate the bit-widths
of the feedforward weights ($fw$), recurrent weights ($rw$),
and neuron parameters: $\beta$ and $\alpha$ for DVS, and
$\beta$ for MNIST. The corresponding color mappings appear
in the legends.

Both axes show weighted costs. The horizontal axis combines
normalized hardware estimates with
$C_{\mathrm{LUT}}=0.2$, $C_{\mathrm{FF}}=0.3$, and
$C_{\mathrm{BRAM}}=0.5$, then applies the hardware weight
$C_{\mathrm{H}}=0.5$. The vertical axis shows classification
error weighted by $C_{\mathrm{A}}=0.5$. Lower values are
preferred on both axes.

The red dashed lines connect the nondominated evaluated
configurations. For the chosen weights, the red circle marks
the evaluated configuration with the smallest sum of the two
plotted costs, as defined in \eqref{eq:total_cost}. The boxed
annotations report the network settings, objective weights,
selected precision configuration, performance, resource
estimates, and search coverage.

The DVS example in Figure~\ref{fig:dse_dvs} uses a
$256$--$128$--$11$ Synaptic network with ATA-T connectivity
and $80$ time steps. Feedforward and recurrent weight
precisions each range from $4$ to $12$ bits, while the
precisions of $\beta$ and $\alpha$ range from $2$ to $7$
bits. This gives $2916$ configurations, of which the search
evaluates $292$ unique configurations ($10.01\%$ coverage).

The selected DVS configuration uses
$(fw,rw,\beta,\alpha)=(9,9,5,3)$ bits and achieves
$87.88\%$ top-1 accuracy. Estimated utilization is
$1757.0$ LUTs and $16.5$ BRAMs. Its weighted hardware and
accuracy costs are approximately $0.2660$ and $0.0606$,
respectively, with a total cost of $0.326609$.

The MNIST example in Figure~\ref{fig:dse_mnist} uses a
$256$--$128$--$10$ LIF network with ATA-T connectivity and
$50$ time steps. Feedforward weight precision ranges from
$4$ to $11$ bits, recurrent weight precision from $4$ to
$12$ bits, and the precision of $\beta$ from $2$ to $7$
bits; $\alpha$ is not applicable. The search evaluates
$216$ of the $432$ configurations ($50\%$ coverage).

The selected MNIST configuration uses
$(fw,rw,\beta)=(6,8,4)$ bits and achieves $92.89\%$ top-1
accuracy, with estimated utilization of $1922.6$ LUTs and
$12.5$ BRAMs. Its weighted hardware and accuracy costs are
approximately $0.2129$ and $0.0355$, respectively, with a
total cost of $0.248499$.

In both examples, configurations with similar hardware costs
can have markedly different accuracy costs. The selected
configurations balance the two terms rather than minimizing
either one alone. Their different precision assignments
also demonstrate the need for application-specific exploration.

\section{Conclusion}\label{sec:conclusion}

This paper presents a comprehensive hardware--software co-design framework for accelerating spiking neural networks (SNNs), centered on two synergistic components: \mbox{Flexi-NeurA} and \mbox{Flex-Plorer}. \mbox{Flexi-NeurA} is a design-time configurable neuromorphic processing core that enables fine-grained parameterization of neuron models, network topologies, reset mechanisms, weight precisions, and synapse capacities. Flexi-NeurA combines event-driven processing with a time-multiplexed
datapath in a synchronous architecture to reduce idle computation
and share arithmetic resources among neurons. Extensive evaluation across three diverse edge benchmarks---spanning auditory classification, event-based vision (DVS gesture recognition), and standard image classification---demonstrates a precise and favorable trade-off between topology complexity and energy efficiency. Complementing the hardware substrate, \mbox{Flex-Plorer} is an automated design-space exploration (DSE) framework that orchestrates parameter configuration, RTL generation, high-level device driver synthesis, and input spike encoding. At its core, it integrates a machine-learning-based hardware cost estimator alongside a dataflow-aware, accuracy simulator, facilitating rapid, low-overhead convergence across large exploration spaces. Overall, this co-design methodology enables highly optimized, application-tailored SNN deployments for resource-constrained edge-AI systems.

\section*{Code and Data Availability}

To encourage research in this field, \mbox{Flex-Plorer} and \mbox{Flexi-NeurA} are available at https://github.com/Mohammadfarahani98.


\begin{thebibliography}{99}

\bibitem{ref1}
R.~Singh and S.~S.~Gill, ``Edge AI: A survey,'' \emph{Internet Things Cyber-Phys. Syst.}, vol.~3, pp.~71--92, 2023.

\bibitem{ref2}
Z.~Zhou \emph{et al.}, ``Edge intelligence: Paving the last mile of artificial intelligence with edge computing,'' \emph{Proc. IEEE}, vol.~107, no.~8, pp.~1738--1762, 2019.

\bibitem{ref3}
W.~Maass, ``Networks of spiking neurons: The third generation of neural network models,'' \emph{Neural Netw.}, vol.~10, no.~9, pp.~1659--1671, 1997.

\bibitem{ref4}
M.~Bouvier \emph{et al.}, ``Spiking neural networks hardware implementations and challenges: A survey,'' \emph{ACM J. Emerg. Technol. Comput. Syst.}, vol.~15, no.~2, pp.~1--35, 2019.

\bibitem{ref5}
K.~Roy, A.~Jaiswal, and P.~Panda, ``Towards spike-based machine intelligence with neuromorphic computing,'' \emph{Nature}, vol.~575, pp.~607--617, 2019.

\bibitem{ref6}
S.~Liu \emph{et al.}, ``An area- and energy-efficient spiking neural network with STDP realized with SRAM PIM macro and on-chip unsupervised learning,'' \emph{IEEE Trans. Biomed. Circuits Syst.}, vol.~17, no.~1, pp.~92--104, 2023.

\bibitem{ref7}
G.~Indiveri and S.-C.~Liu, ``Memory and information processing in neuromorphic systems,'' \emph{Proc. IEEE}, vol.~103, no.~8, pp.~1379--1397, 2015.

\bibitem{ref8}
A.~Carpegna, A.~Savino, and S.~D.~Carlo, ``Spiker+: A framework for the generation of efficient spiking neural networks FPGA accelerators for inference at the edge,'' \emph{IEEE Trans. Emerg. Topics Comput.}, 2024.

\bibitem{ref9}
C.~Frenkel, M.~Lefebvre, J.-D.~Legat, and D.~Bol, ``A 0.086-mm$^{2}$ 12.7-pJ/SOP 64k-synapse 256-neuron online-learning digital spiking neuromorphic processor in 28-nm CMOS,'' \emph{IEEE Trans. Biomed. Circuits Syst.}, vol.~13, 2019.

\bibitem{ref10}
C.~Frenkel and G.~Indiveri, ``ReckOn: A 28-nm sub-mm$^{2}$ task-agnostic spiking recurrent neural network processor enabling on-chip learning over second-long timescales,'' in \emph{Proc. IEEE Int. Solid-State Circuits Conf. (ISSCC)}, 2022.

\bibitem{ref11}
C.~Frenkel, J.-D.~Legat, and D.~Bol, ``MorphIC: A 65-nm 738k-synapse/mm$^{2}$ quad-core binary-weight digital neuromorphic processor with stochastic spike-driven online learning,'' \emph{IEEE Trans. Biomed. Circuits Syst.}, pp.~1--11, 2019.

\bibitem{ref12}
D.~Padovano, A.~Carpegna, A.~Savino, and S.~D.~Carlo, ``SpikeExplorer: Hardware-oriented design space exploration for spiking neural networks on FPGA,'' \emph{Electronics}, vol.~13, no.~9, May~2024.

\bibitem{ref13}
N.~Jannesar, K.~Akbarzadeh-Sherbaf, S.~Safari, and A.-H.~Vahabie, ``SSTE: Syllable-specific temporal encoding to FORCE-learn audio sequences with an associative memory approach,'' \emph{Neural Netw.}, vol.~177, 2024.

\bibitem{ref14}
J.~K.~Eshraghian \emph{et al.}, ``Training spiking neural networks using lessons from deep learning,'' \emph{Proc. IEEE}, vol.~111, no.~9, pp.~1016--1045, 2023.

\bibitem{ref15}
A.~L.~Hodgkin and A.~F.~Huxley, ``A quantitative description of membrane current and its application to conduction and excitation in nerve,'' \emph{J. Physiol.}, vol.~117, no.~1, pp.~500--544, 1952.

\bibitem{ref16}
K.~Akbarzadeh-Sherbaf, B.~Abdoli, S.~Safari, and A.-H.~Vahabie, ``A scalable FPGA architecture for randomly connected networks of Hodgkin--Huxley neurons,'' \emph{Front. Neurosci.}, vol.~12, 2018.

\bibitem{ref17}
S.~Y.~Bonabi, H.~Asgharian, S.~Safari, and M.~N.~Ahmadabadi, ``FPGA implementation of a biological neural network based on the Hodgkin--Huxley neuron model,'' \emph{Front. Neurosci.}, vol.~8, 2014.

\bibitem{ref18}
D.-A.~Nguyen, X.-T.~Tran, and F.~Iacopi, ``A review of algorithms and hardware implementations for spiking neural networks,'' \emph{J. Low Power Electron. Appl.}, vol.~11, no.~2, 2021.

\bibitem{ref19}
F.~Akopyan \emph{et al.}, ``TrueNorth: Design and tool flow of a 65~mW 1 million neuron programmable neurosynaptic chip,'' \emph{IEEE Trans. Comput.-Aided Des. Integr. Circuits Syst.}, vol.~34, no.~10, pp.~1537--1557, Oct.~2015.

\bibitem{ref20}
M.~Sadeghi \emph{et al.}, ``NEXUS: A 28~nm 3.3~pJ/SOP 16-core spiking neural network with a diamond topology for real-time data processing,'' \emph{IEEE Trans. Biomed. Circuits Syst.}, vol.~19, no.~3, pp.~523--535, Jun.~2025.

\bibitem{ref21}
J.~Han, Z.~Li, W.~Zheng, and Y.~Zhang, ``Hardware implementation of spiking neural networks on FPGA,'' \emph{Tsinghua Sci. Technol.}, vol.~25, no.~4, pp.~479--486, 2020.

\bibitem{ref22}
S.~Li \emph{et al.}, ``A fast and energy-efficient SNN processor with adaptive clock/event-driven computation scheme and online learning,'' \emph{IEEE Trans. Circuits Syst.~I: Reg. Papers}, vol.~68, no.~4, pp.~1543--1552, 2021.

\bibitem{ref23}
S.~K.~Lam, A.~Pitrou, and S.~Seibert, ``Numba: A LLVM-based Python JIT compiler,'' in \emph{Proc. 2nd Workshop LLVM Compiler Infrastructure in HPC (LLVM'15)}, 2015.

\bibitem{ref24}
S.~Gupta, A.~Vyas, and G.~Trivedi, ``FPGA implementation of simplified spiking neural network,'' in \emph{Proc. 27th IEEE Int. Conf. Electron., Circuits Syst.}, 2020.

\bibitem{ref25}
A.~Carpegna, A.~Savino, and S.~D.~Carlo, ``Spiker: An FPGA-optimized hardware accelerator for spiking neural networks,'' in \emph{Proc. IEEE Comput. Soc. Annu. Symp. VLSI (ISVLSI)}, 2022, pp.~14--19.

\bibitem{ref26}
D.~Kudithipudi \emph{et al.}, ``Neuromorphic computing at scale,'' \emph{Nature}, vol.~637, no.~8047, pp.~801--812, 2025.

\bibitem{ref27}
E.~Covi \emph{et al.}, ``Adaptive extreme edge computing for wearable devices,'' \emph{Front. Neurosci.}, vol.~15, p.~611300, 2021.

\bibitem{ref28}
A.~Amir \emph{et al.}, ``A low power, fully event-based gesture recognition system,'' in \emph{Proc. IEEE Conf. Comput. Vis. Pattern Recognit. (CVPR)}, 2017, pp.~7243--7252.


\bibitem{ref29}
Y.~LeCun and C.~Cortes, ``The MNIST database of handwritten digits,'' 1998. [Online]. Available: \url{http://yann.lecun.com/exdb/mnist/}. Accessed: May~2026.

\bibitem{ref30}
B.~Cramer \emph{et al.}, ``The Heidelberg spiking data sets for the systematic evaluation of spiking neural networks,'' \emph{IEEE Trans. Neural Netw. Learn. Syst.}, vol.~33, no.~7, pp.~2744--2757, 2022.

\bibitem{ref31}
M.~R.~Roshanshah, S.~Safari, and M.~Modarressi, ``ChronoSpike: A low-cost, deterministic circuit-switched NoC for neuromorphic accelerators,'' \emph{Neurocomputing}, vol.~699, Art.~no.~134364, 2026.

\bibitem{ref32}
A.~Alipour~Fereidani, M.~R.~Roshanshah, and S.~Safari, ``ReSCom: A reconfigurable spiking neural network accelerator using stochastic computing,'' \emph{Integration}, vol.~112, Art.~no.~102906, 2027.

\bibitem{ref33}
K.~Gunawardana, S.~Peeris, K.~Rambukwella, R.~Ragel, and I.~Nawinne, ``Neuromorphic architectures for edge-oriented spiking neural networks: A review,'' \emph{J. Syst. Archit.}, vol.~177, p.~103869, 2026.

\bibitem{ref34}
J.~Wang, Y.~Shi, Z.~Chen, and M.~Wen, ``SpikeFlow: A hardware--software co-designed systolic array for spiking neural networks,'' \emph{J. Syst. Archit.}, vol.~169, p.~103604, 2025.

\bibitem{ref35}
C.~Xiao, Y.~Wei, Y.~Luo, Z.~Yang, R.~Gong, W.~Xu, and L.~Wang, ``Hardware/software co-design for spike communication optimization: Leveraging neuron-level communication patterns,'' \emph{J. Syst. Archit.}, vol.~168, p.~103553, 2025.

\bibitem{ref36}
A.~Parashar \emph{et al.}, ``Timeloop: A systematic approach to DNN accelerator evaluation,'' in \emph{Proc. IEEE Int. Symp. Perform. Anal. Syst. Softw. (ISPASS)}, 2019, pp.~304--315.

\bibitem{ref37}
M.~Davies \emph{et al.}, ``Loihi: A neuromorphic manycore processor with on-chip learning,'' \emph{IEEE Micro}, vol.~38, no.~1, pp.~82--99, 2018.

\bibitem{ref38}
S.~S.~Ghavami, M.~H.~Nikkhah, M.~R.~Roshanshah, and S.~Safari, ``SupraSNN: Exploiting Synapse-Level Parallelism in Spiking Neural Network Accelerators through Co-Optimized Mapping and Scheduling,'' arXiv preprint arXiv:2606.13354, 2026. [Online]. Available: \url{https://arxiv.org/abs/2606.13354}



\end{thebibliography}
\end{document}